\documentclass[twocolumn, twocolappendix]{aastex631}

\hypersetup{linkcolor=blue,citecolor=blue,filecolor=cyan,urlcolor=cyan}

\newcommand{\mr}[1]{\mathrm{#1}}
\newcommand{\kmps}{\mr{km~s^{-1}}}
\newcommand{\VT}{Department of Physics, Virginia Tech, 850 West Campus Drive, Blacksburg, VA 24061, USA}
\newcommand{\UH}{Institute for Astronomy, University of Hawai'i at Manoa, 2680 Woodlawn Dr., Hawai'i, HI 96822, USA }

\begin{document}

\title{Using nebular near-IR spectroscopy to measure asymmetric chemical distributions in 2003fg-like thermonuclear supernovae}

\author[0009-0007-5201-7954]{J. O'Hora}
\affiliation{\VT}

\author[0000-0002-5221-7557]{C. Ashall} 
\correspondingauthor{Chris Ashall}
\email{cashall@hawaii.edu}
\affiliation{\VT}
\affiliation{\UH}

\author[0000-0002-9301-5302]{M. Shahbandeh}
\affiliation{Space Telescope Science Institute, 3700 San Martin Drive, Baltimore, MD 21218-2410, USA}

\author[0000-0003-1039-2928]{E. Hsiao}
\affiliation{Department of Physics, Florida State University, 77 Chieftan Way, Tallahassee, FL 32306, USA}

\author[0000-0002-4338-6586]{P. Hoeflich}
\affiliation{Department of Physics, Florida State University, 77 Chieftan Way, Tallahassee, FL 32306, USA}

\author[0000-0002-5571-1833]{M. D. Stritzinger}
\affiliation{Department of Physics and Astronomy, Aarhus University, Ny Munkegade 120, DK-8000 Aarhus C, Denmark}

\author[0000-0002-1296-6887]{L. Galbany}
\affiliation{Institute of Space Sciences (ICE, CSIC), Campus UAB, Carrer de Can Magrans, s/n, E-08193 Barcelona, Spain}
\affiliation{Institut d’Estudis Espacials de Catalunya (IEEC), E-08034 Barcelona, Spain}

\author[0000-0001-5393-1608]{E. Baron}
\affiliation{Planetary Science Institute, 1700 East Fort Lowell Road, Suite 106,
 Tucson, AZ 85719-2395 USA}
\affiliation{Hamburger Sternwarte, Gojenbergsweg 112, D-21029 Hamburg, Germany}

\author[0000-0002-7566-6080]{J. DerKacy}
\affiliation{Space Telescope Science Institute, 3700 San Martin Drive, Baltimore, MD 21218-2410, USA}
\affiliation{\VT}

\author[0000-0001-8367-7591]{S. Kumar}
\affiliation{Department of Astronomy, University of Virginia, 530 McCormick Rd, Charlottesville, VA 22904, USA}

\author[0000-0002-3900-1452]{J. Lu}
\affil{Department of Physics and Astronomy, Michigan State University, East Lansing, MI 48824, USA}

\author[0000-0001-7186-105X]{K. Medler}
\affiliation{\VT}
\affiliation{\UH}

\author[0000-0003-4631-1149]{B. Shappee}
\affiliation{\UH}

\begin{abstract} 
\noindent We present an analysis of three near-infrared (NIR; 1.0-2.4$~\micron$) spectra of the SN~2003fg-like/``super-Chandrasekhar" type Ia supernovae (SNe~Ia) SN~2009dc, SN~2020hvf, and SN~2022pul at respective phases of +372, +296, and +294~d relative to the epoch of $B$-band maximum. We find that all objects in our sample have asymmetric, or ``tilted," [Fe~II] 1.257 and 1.644~$\micron$ profiles.
We quantify the asymmetry of these features using five methods: velocity at peak flux, profile tilts, residual testing, velocity fitting, and comparison to deflagration-detonation transition models. 
Our results demonstrate that, while the profiles of the [Fe~II] 1.257 and 1.644~$\micron$ features are widely varied between 2003fg-likes, these features are correlated in shape within the same SN. This implies that line blending is most likely not the dominant cause of the asymmetries inferred from these profiles. Instead, it is more plausible that 2003fg-like SNe have aspherical chemical distributions in their inner regions. These distributions may come from aspherical progenitor systems, such as double white dwarf mergers, or off-center delayed-detonation explosions of near-Chandrasekhar mass carbon-oxygen white dwarfs. 
Additional late-phase NIR observation of 2003fg-like SNe and detailed 3-D NLTE modeling of these  two explosion scenarios are encouraged.
\end{abstract}


\section{Introduction} \label{sec:intro}
Type Ia supernovae (SNe~Ia) are thermonuclear explosions of carbon/oxygen (C/O) white dwarf (WD) stars in binary systems~\citep{HoyleFowler_1960_sneia,HillebrandtNiemeyer_2000_sneia}. Despite decades of detailed study, the exact nature of their progenitors and explosion mechanisms are a matter of open debate \citep{IbenTutukov_1984_binaries_gws,Webbink_1984_doublewd,Fink_2007_doubledet,Mazzali_2007_nebphase,Wang_2012_iaprogenitors,Holcomb_2013_hedet,Liu_2017_cowd_herichwd,Liu_2023_iaexplosions}.  It has long been apparent that SNe~Ia are a diverse group comprised of a variety of ``peculiar" sub-classes~\citep[e.g.][]{Filippenko_1992_phot_91bg,Filippenko_1992_91t,Mazzali_1995_91t,Howell_2001_91bg,Li_2001_peculrate,Li_2003_02cx,Ganeshalingam_2012_02es}. 
Some of the SNe~Ia within these sub-classes do not follow the luminosity-width relation, suggesting that there are multiple pathways to produce these cosmic explosions \citep[e.g.][]{Livio18,Ashall_2021_03fglikes,Liu_2023_iaexplosions,Siebert_2023_22pul}.

One rare sub-class are 2003fg-like/``super-Chandrasekhar'' SNe (hereafter ``03fg-likes";~\citealt{Howell_2006_03fg}). 
These objects generally exhibit: \textit{i)} high peak absolute $B$-band magnitudes spanning the range between $\mr{-21<M_{B}<-19}$~mag; \textit{ii)} broad light curves with decline rates of $\mr{\Delta m_{\mr{15},B}}<1.3$~mag; \textit{iii)} strong C absorption at maximum light; \textit{iv)} low ejecta velocities; \textit{v)} an $i$-band light curve peak that occurs after the $B$-band peak; \textit{vi)} higher luminosities in the near-infrared (NIR) compared to normal SNe~Ia; \textit{vii)} a tendency to occur in low-metallicity, low-surface-brightness galaxies; and \textit{viii)} distinct UV colors~(\citealt{Howell_2006_03fg,Childress_2011_metalpoor_07if,Khan_2011_metalpoor,Taubenberger_2019_12dn,Ashall_2021_03fglikes,Liu_2023_iaexplosions,Hoogendam_2024_03fgUV}; Galbany et al. in prep). 
Some 03fg-like SNe~Ia also exhibit an early flux excess (or a ``bump") in their light curves relative to a smooth power-law rise~\citep[e.g.][]{Jiang_2021_20hvf,Dimitriadis_2023_21znydd,Siebert_2023_asymmetricdd,Srivastav_2023_22ilv}.

The precise workings of the progenitor systems and explosion mechanisms of 03fg-like SNe are debated. However, it is generally accepted that these explosions occur within a carbon-rich  envelope ~\citep{Hachinger_2012_coredegen,Noebauer_2016_csm,Ashall_2021_03fglikes,Maeda_2023_envelope_overluminous}. There are two popular models for explaining this carbon-rich ejecta: textit{i)} the violent merger of two C/O WDs~\citep{Webbink_1984_doublewd,Fink_2007_doubledet,Pakmor_2010_subluminousmergers,Pakmor_2011_violentsubluminous,Pakmor_2012_violentmergers,Siebert_2023_22pul,Siebert_2023_asymmetricdd,Srivastav_2023_22ilv,Kwok_2024_22pul}, which is naturally asymmetric; or \textit{ii)} the core-degenerate scenario, where a WD merges with the degenerate core of an asymptotic giant branch star~\citep{Livio_2003_iaprogenitors,KashiSoker_2011_coredegen,Hsiao_2020_carnegieii,Lu_2021_15hy}.
In the case of the latter, a thermonuclear runaway may occur upon the initiation of a subsonic deflagration flame front, which may then transition to a supersonic detonation flame ~\citep{Hoeflich_2019_multiwavelength,Soker_2019_spherical}. 
In this scenario, the deflagration-to-detonation transition (DDT) occurs 
on an approximately-spherical expanding background, which results in lower density burningd thus asymmetric abundance distributions of
burning products produced during the detonation \citep[e.g.][]{Hoeflich_2021_20qxp,Derkacy_2023_21aefx,Ashall24_JWST}.

Overall, both explosion scenarios are capable of producing the high luminosity and C I and C II absorption observed in 03fg-like SNe~\citep[e.g.][]{Lu_2021_15hy,Siebert_2023_22pul,Kwok_2024_22pul}; however, each scenario has caveats. 
In the core-degenerate scenario, it is not clear how the progenitor system would shed its H and He-rich envelope. It is plausible that this could occur through a super-wind phase~\citep{Hsiao_2020_carnegieii}. In the violent merger scenario, on the other hand, it is not clear how this scenario could incorporate a massive C-rich envelope, and further work is also needed to determine if these systems can produce dust \citep[e.g.][]{Siebert_2023_22pul,Kwok_2024_22pul,Liu_2025_CIin22pul}.

Distinguishing between these two progenitor models is not trivial. This is because both scenarios may have an asymmetric chemical distribution in the core of the explosion, which is highly dependent on the triggering mechanism of the explosion and how the flame propagates through the ejecta. Recently, \citet{Liu_2025_CIin22pul}  used optical spectra to infer further information about the explosion mechanism of these events, suggesting that 
centrally-peaked [C~I] 9.824 and 9.850~$\micron$ emissions in
the 03fg-like SN~2022pul at a phase of +515~d from $B$-band maximum may indicate a pure deflagration of a super-Chandrasekhar-mass WD \citep{Liu_2025_CIin22pul}.

In this work, our attention is focused on late-phase observations in the NIR. Nebular-phase ($>$250~d) spectra of SNe~Ia reveal the inner region of the ejecta, making them a useful resource for analyzing core asymmetry, burning conditions, and explosion physics~\citep{Marietta_2000_neb_progenitor,Mazzali_2007_nebphase,Maeda_2010_asymmetries,Hoeflich_2021_20qxp,kumar_2023_13aa17cbv,Dimitriadis_2023_21znydd,kumar_2023_13aa17cbv,Ashall24_JWST,Bose25,Liu_2025_CIin22pul}. 
Furthermore, the NIR is a useful yet underutilized wavelength region compared to the optical, since it provides a different set of spectral features and exhibits less line blending. For example, the [Fe~II] emission lines at 1.257 and 1.644~$\micron$ have been shown to provide measurements of explosion kinematics, WD central density, and magnetic fields in normal SNe~Ia~\citep{Penney_2014_bfields,Diamond_2015_nir,Diamond_2018_14j_164,Maguire_2018_126,Hoeflich_2021_20qxp,kumar_2023_13aa17cbv}. To date, SN~2009dc and SN~2022pul are the only 03fg-like SNe with a published nebular-phase-NIR spectrum~\citep{taubenberger_2013_nebular03fg,Siebert_2023_22pul,Kwok_2024_22pul}. Moreover, no work thus far has analyzed a sample of 03fg-likes in both the NIR and at nebular phases to look for similarities between the explosions. 

Here, we analyze NIR nebular spectra of SN~2009dc and SN~2022pul in conjunction with a new NIR nebular spectrum of the 03fg-like SN~2020hvf. In Section~\ref{sec:data_reduction}, we present details of the observations, as well as data reduction when appropriate. In Section~\ref{sec:line_ids}, we discuss the lines that contribute to the spectra, as well as the shapes of the spectral features. In Section~\ref{sec:analytical}, we compare the spectroscopy of  03fg-like SNe to each other and normal SNe~Ia, with specific emphasis on the asymmetric line profiles of the 03fg-likes. 
Section~\ref{sec:analytical} also presents five analytical methods to quantitatively describe these profile asymmetries: Section~\ref{subsec:analytical_peaks} presents our measurements of the velocities at peak flux for the 03fg-like and normal SNe; Section~\ref{subsec:analytical_tilts} quantifies the degree of asymmetry/tilting in the 03fg-like features; Section~\ref{subsec:analytical_residuals} presents a residual test between the [Fe~II] 1.644$~\mr{\micron}$ lines of each 03fg-like with each normal SN Ia; Section~\ref{subsec:analytical_gaussian} discusses a curve-fitting procedure for the 03fg-like spectra that visualizes the individual components of the complexes around 1.26 and 1.64$~\mr{\micron}$ with Gaussian functions; and lastly, in Section~\ref{subsec:analytical_ddt} we compare the 1.644~$\micron$ features of the 03fg-likes to pre-existing spectral models of off-center delayed-detonation explosions. Our conclusions are presented in Section~\ref{sec:concl}.

\begin{deluxetable*}{lccllcccc}[t]
  \tablecaption{Log of observations used in this work. \label{tab:obs}} 
  \tablehead{\colhead{SN} & \colhead{RA (J2000)} & \colhead{Dec (J2000)}& \colhead{Subtype}& \colhead{z}&\colhead{Phase\tablenotemark{a}}&\colhead{Telescope+Instrument}&\colhead{Resolution}&\colhead{Reference}}
  \startdata
    2009dc&15:51:12.12 &+25:42:28.00&03fg-like&0.0214&+372&VLT+XShooter&5573&\textit{b}\\
    2013aa&14:32:33.88 &$-$44:13:27.80&Normal&0.0040\tablenotemark{c}&+365&Magellan Baade+FIRE&1200&\textit{d}\\
    2017cbv&14:32:34.42 &$-$44:08:02.74&Normal&0.0040\tablenotemark{c}&+307&Magellan Baade+FIRE&1200&\textit{d}\\
    2020hvf&11:21:26.45 &+03:00:52.85&03fg-like&0.0058&+296&Keck II+NIRES&2700&This work\\
    2022pul&12:26:49.00 &+08:26:55.25&03fg-like&0.0030&+294&Keck II+NIRES&2700&\textit{e}\\
  \enddata
  \tablecomments{\tablenotemark{a} Rest-frame days relative to the epoch of  $B$-band maximum; \tablenotemark{b}~\cite{taubenberger_2013_nebular03fg}; \tablenotemark{c} Additional corrections are made for the motion of the SN within the galaxy (see~\citealt{kumar_2023_13aa17cbv} for more details); \tablenotemark{d}~\cite{kumar_2023_13aa17cbv}; \tablenotemark{e}~\cite{Siebert_2023_22pul}.}
\end{deluxetable*}

\section{Observations \& Data Reduction} \label{sec:data_reduction} 

For this work we searched the literature for NIR spectra of 03fg-like SNe in the nebular phase ($>$+250~d).
To our knowledge, only SN~2009dc and SN~2022pul have such published observations. We analyze these spectra along with a new NIR nebular phase spectrum of SN~2020hvf, giving us a total of three nebular-NIR spectra of 03fg-like SNe. To emphasize the differences in spectral profiles between 03fg-like and normal SNe~Ia, we also analyze NIR nebular phase spectra of SNe~Ia~2013aa and 2017cbv \citep{kumar_2023_13aa17cbv}.  We choose these objects because they are well observed and their nebular phase spectra are representative of normal SNe~Ia, which generally exhibit highly symmetric [Fe~II] line profiles~(\citealt{Diamond_2015_nir,Diamond_2018_14j_164,kumar_2023_13aa17cbv,Kumar24}, Kumar et al., in preparation). Table~\ref{tab:obs} presents the full log of observations used throughout this work. 

The NIR spectrum of SN~2020hvf was obtained using the Near-Infrared Echellette Spectrometer (NIRES) instrument on the Keck~II telescope on 2021 March 4.40 UT ($\mr{MJD~59277.4}$). This date corresponds to a rest-frame phase of +296.4~d from the epoch of $B$-band maximum which occurred on $\mr{MJD~58979.3}$~\citep{Jiang_2021_20hvf}. The spectrum covers the wavelength region between $\mr{0.97\mu m\sim2.47\mu m}$. The total exposure time of the observations was 7200~s from 6 sets of ABBA observations, where each individual A/B exposure lasted 300~s. The data was reduced with the IDL package Spextools version 5.0.2~\citep{Cushing_2004_Spextool}. Flux calibration and telluric feature corrections were completed with Xtellcorr version 5.0.2 via observations of the A0V star HIP56736. Throughout this work, spectral resolution has been accounted for when reporting velocity measurements. 

\begin{figure*} 
    \centering
    \includegraphics[width=\textwidth]{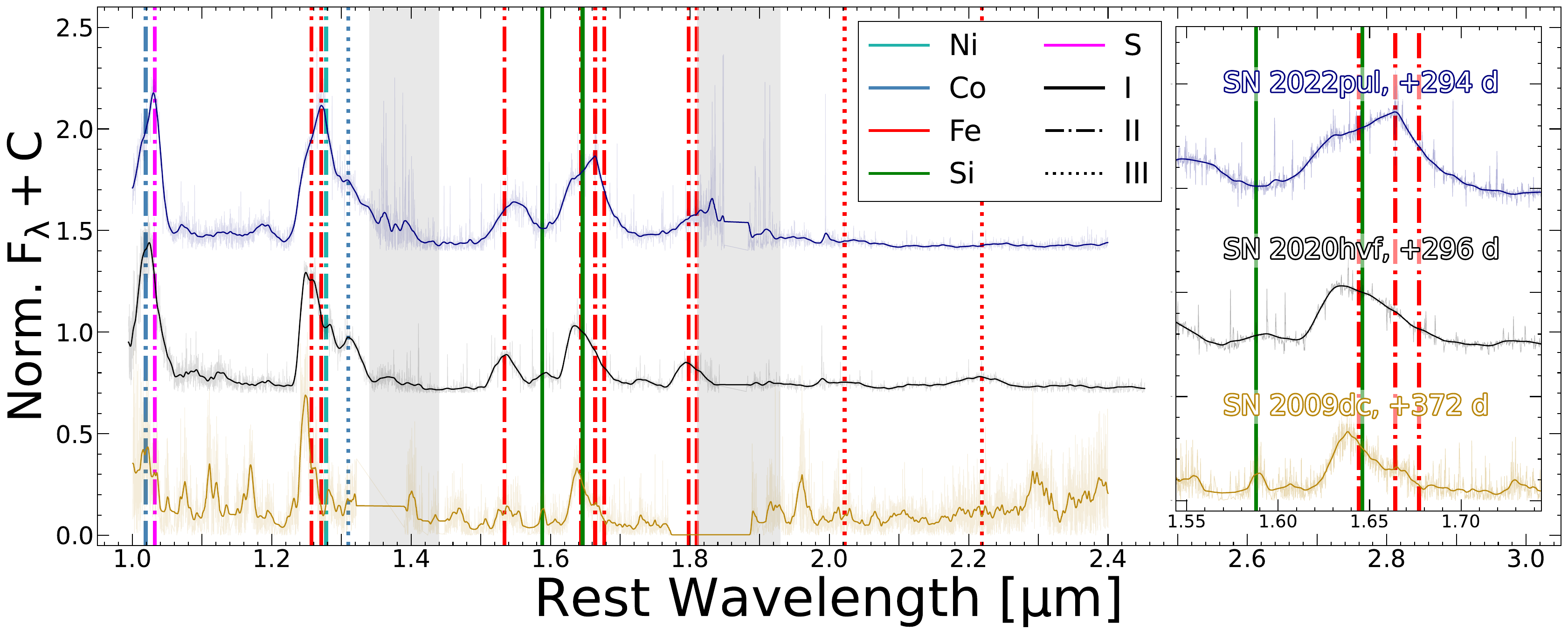}
    \caption{Full NIR spectra of SN~2009dc, SN~2020hvf, and SN~2022pul at nebular phases. Spectra are redshift-corrected (Table~\ref{tab:obs}), offset from each other, and overlaid with a smooth 1D Gaussian-filtered solid curve with a smoothing length of 12 pixels. Prominent spectral features are marked, with color- and style-coding by ionic species. Telluric regions are highlighted in light gray. The inset identifies many lines that could contribute to the feature around 1.6~$\micron$ for each 03fg-like SN presented. However, as discussed in the text (Section~\ref{subsec:analytical_tilts}), we identify  [Fe II] 1.644~$\micron$ as the dominant ion in this region.}
    \label{fig:spectrum_03fglikes_full}
\end{figure*}

\begin{deluxetable*}{cccccccccc}[t] 
  \tablecaption{List of identified spectral lines \label{tab:lineids}} 
  \tablehead{\colhead{$\lambda\mr{~(\micron)}$} & \colhead{Ion} & \colhead{$\lambda\mr{~(\micron)}$} & \colhead{Ion} & \colhead{$\lambda\mr{~(\micron)}$} & \colhead{Ion} & \colhead{$\lambda\mr{~(\micron)}$} & \colhead{Ion} & \colhead{$\lambda\mr{~(\micron)}$} & \colhead{Ion}}
  \startdata
    0.994 & [Co~II] & 1.046 & [Ni~II] & 1.298 & [S~III] & 1.547 & [Co~II] & 1.725 & [Ni~II] \\
    1.019 & [Co~II] & 1.131 & [Si~I]  & 1.310 & [Co~III]& 1.588 & [Si~I]  & 1.741 & [Fe~II] \\
    1.021 & [Ni~II] & 1.191 & [Fe~II] & 1.321 & [Fe~II] & 1.644 & [Fe~II] & 1.745 & [Fe~II] \\
    1.024 & [Co~II] & 1.257 & [Fe~II] & 1.328 & [Fe~II] & 1.646 & [Si~I]  & 1.798 & [Fe~II] \\
    1.028 & [Co~II] & 1.271 & [Fe~II] & 1.372 & [Fe~II] & 1.664 & [Fe~II] & 1.810 & [Fe~II] \\
    1.029 & [S~II]  & 1.278 & [Ni~II] & 1.525 & [Fe~II] & 1.677 & [Fe~II] & 2.022 & [Fe~III]\\
    1.032 & [S~II]  & 1.294 & [Fe~II] & 1.534 & [Fe~II] & 1.712 & [Fe~II] & 2.219 & [Fe~III]\\
  \enddata
  \tablecomments{Line identifications are chosen based on radiation transport models from~\cite{Hoeflich_2021_20qxp} and~\cite{Blondin_2023_21aefx_models}.}
\end{deluxetable*}

\section{Line Identifications} 
\label{sec:line_ids}

Figure~\ref{fig:spectrum_03fglikes_full} presents the NIR spectra of SNe~2009dc, 2020hvf, and 2022pul at respective phases of +372~d, +296~d, and +294~d from $B$-band maximum, which we adopt in this work as the standard reference for phase. Using the line lists from~\cite{Hoeflich_2021_20qxp},~\cite{ Blondin_2023_21aefx_models}, and~\cite{Kwok_2024_22pul}, we identify the strongest lines that may contribute to each spectral feature. 
Our line IDs are listed in Table~\ref{tab:lineids}. We discuss these lines from the bluer to the redder wavelengths below. Future modeling is required to ascertain exactly which lines contribute to the spectral formation and what strengths these lines have.

The strong peak at the bluest ends of all spectra can be attributed to a blend of [Co~II] 1.019 and [S~II] 1.032~$\micron$. This complex is difficult to identify in SN~2009dc due to noise. For all SNe, a blended complex spans the wavelength region between 1.24-1.34~$\micron$. The bluer end of this complex is dominated by emission associated with [Fe~II] 1.257~$\micron$. Weak [Ni~II] 1.278~$\micron$ is observed in all three SNe just redwards of the 1.257~$\micron$ line; we also observe weak [Co~III] 1.310~$\micron$ in SN~2020hvf and SN~2022pul only. Moving redward, there exists weak [Fe~II] 1.534~$\micron$ and [Si~I] 1.588~$\micron$ emissions. Strong [Fe~II] 1.644~$\micron$ emission is observed in all SNe. Finally, very weak [Fe~III] 2.022 and 2.219~$\micron$ emissions are observed at the reddest ends of all three spectra. Overall, the weakness of higher-ionization lines (namely [Fe~III]) in tandem with the strengths of singly-ionized features from the same elements suggest a low ionization state for the 03fg-like SNe relative to normal SNe~Ia. This aligns closely with optical observations of other 03fg-like SNe at these phases~\citep{Silverman_2011_09dc,taubenberger_2013_nebular03fg,Taubenberger_2019_12dn,Ashall_2021_03fglikes}. 

While it may be possible that the line profiles at 1.257 and 1.644~$\micron$ have some contribution from [Fe~II] 1.271~$\micron$, [Si~I] 1.646~$\micron$, [Fe~II] 1.664~$\micron$, and/or [Fe~II] 1.667~$\micron$ we determine this is unlikely. This is because these emissions are expected to be very weak in comparison to [Fe~II] 1.257 and 1.644~$\micron$ \citep{Diamond_2015_nir,Hoeflich_2021_20qxp} and, if these lines were in fact dominant, there would not be similarity in profile shape in the 1.257 and 1.644~$\micron$ regions (see Section~\ref{sec:analytical}).

\section{Spectral Analysis} \label{sec:analytical}

\subsection{Spectral Profiles} \label{sec:spec_profs}
\begin{figure*}[t]
    \centering
    \begin{minipage}{0.47\textwidth}
        \includegraphics[width=\linewidth]{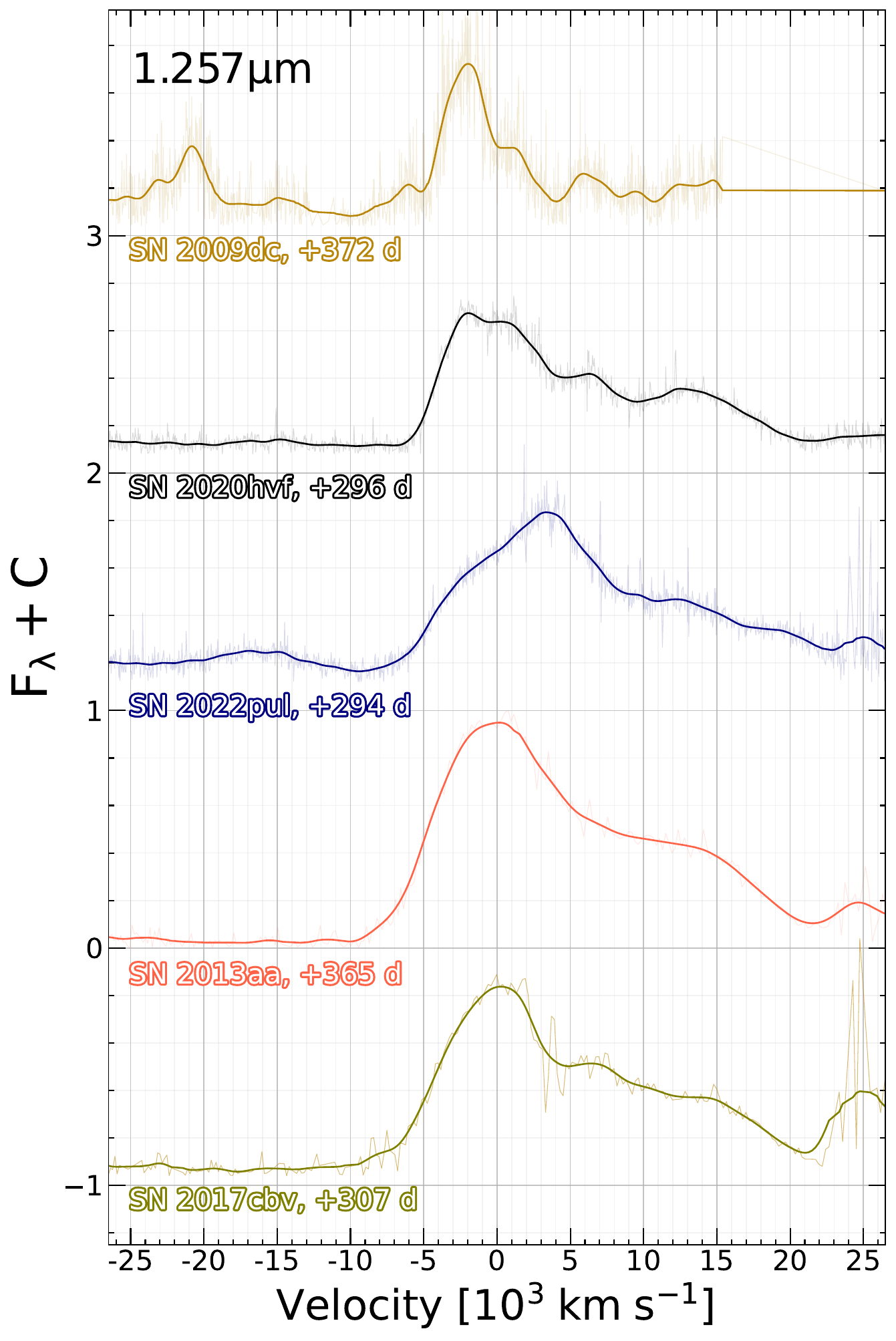}
        \label{fig:plot1}
    \end{minipage}\hfill
    \begin{minipage}{0.485\textwidth}
        \includegraphics[width=\linewidth]{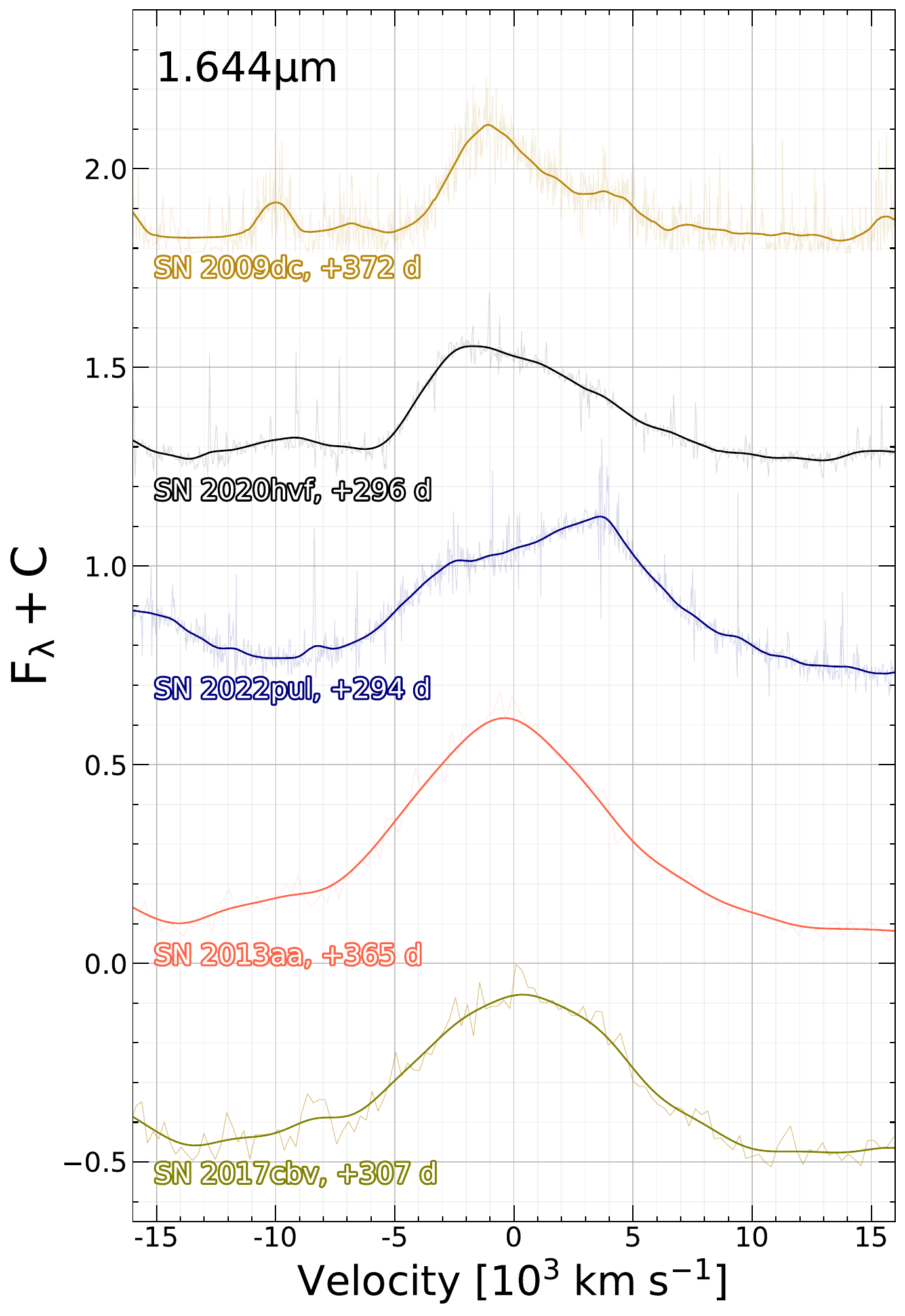}
        \label{fig:plot2}
    \end{minipage}
    \caption{The spectra presented in Table \ref{tab:obs} zoomed in on the [Fe~II] 1.257 (\textit{left}) and 1.644~$\micron$ (\textit{right}) features and plotted in velocity space. Again, spectra are redshift-corrected, offset, and overlaid with a 1D Gaussian-filtered smooth curve.}
    \label{fig:spec_compare_all_feii}
\end{figure*}
As mentioned before, the [Fe~II] 1.257 and 1.644~$\micron$ lines have been shown to be useful in quantitative analyses of SNe Ia~\citep{Penney_2014_bfields,Diamond_2015_nir,Diamond_2018_14j_164,Maguire_2018_126,Hoeflich_2021_20qxp,kumar_2023_13aa17cbv}. For normal SNe Ia, these lines typically exhibit a high degree of symmetry.

In the 03fg-like SNe, however, both features around the [Fe~II] 1.26 and 1.64~$\micron$ regions show a clear asymmetry, appearing as a ``tilt" in their peaks. Furthermore, for each SN, the directions and morphologies of these tilts appear to be consistent between the 1.257 and 1.644~$\micron$ features (see Fig.~\ref{fig:spec_compare_all_feii}). 
For this similarity in tilting between features to be caused by line blending would be an unlikely case; this suggests that [Fe II] is the dominant ion contributing to the spectral formation in both features, in particular [Fe II] 1.257 and 1.644~$\micron$ emissions.

To better qualitatively compare the differences in the [Fe~II] 1.257 and 1.644~$\micron$ features amongst our SNe, Fig.~\ref{fig:spec_compare_all_feii} presents these features in velocity space for each 03fg-like SN in our sample along with SNe~2013aa (+365~d) and 2017cbv (+307~d). The phases of these spectra are each representative of similar late-phase ionization states for SNe~Ia, hence providing solid ground for spectral comparison. For the normal SNe, the peaks of the [Fe~II] 1.257 and 1.644~$\micron$ lines are very close to zero velocity; however, in the 03fg-like SNe, the peaks of these features are largely offset from zero velocity by up to $\sim\pm3000~\mr{km~s^{-1}}$. Additionally, the shapes of the line profiles vary between SNe; SNe~2009dc and 2020hvf exhibit blue-peaked tilts, while SN~2022pul exhibits red-peaked tilts.

\subsection{Profile Peaks} \label{subsec:analytical_peaks}
To better understand the shape of the spectral profiles, we first 
measure the velocity at peak flux of the features without making any assumptions about the shape of the emitting region or the number of different components and ions which contribute to it. This is because the emitting region could be spherical, asymmetric, or ring-like and still be produced from emission dominated by a single ion \citep[e.g.][]{Mazzali_2007_nebphase,Jerkstrand17,Hoeflich_2021_20qxp,Ashall24_JWST}. 
For example, the merger of two WDs may produce a ring-like structure and double-horned profiles~\citep{Jerkstrand17,Hoeflich_2021_20qxp}. Moreover, the suppression of the red component of a spectral feature could also result from the obstruction of flux originating from an extended photosphere \citep{Penney14}.


To obtain measurements of the velocity at peak flux, we  smooth the data using a 1-D Gaussian filter with a smoothing length of 10 pixels. The residual between an individual raw data point $F_i$ and this Gaussian-smoothed data is treated as one standard deviation of error $\sigma_i$. For 500 iterations, a new ``perturbed" smooth curve is then constructed from the original smooth curve by varying each $F_i$ by an amount randomly sampled from a normal distribution with standard deviation $\sigma_i$. The velocity at the absolute maximum of this perturbed smooth curve is then measured for each iteration, giving us a sample of 500 peak velocity values. The mean and standard deviation of this sample thus constitutes our measurement of the velocity at the peak flux of the line.

\begin{figure}[h]
    \centering
    \includegraphics[width=\linewidth]{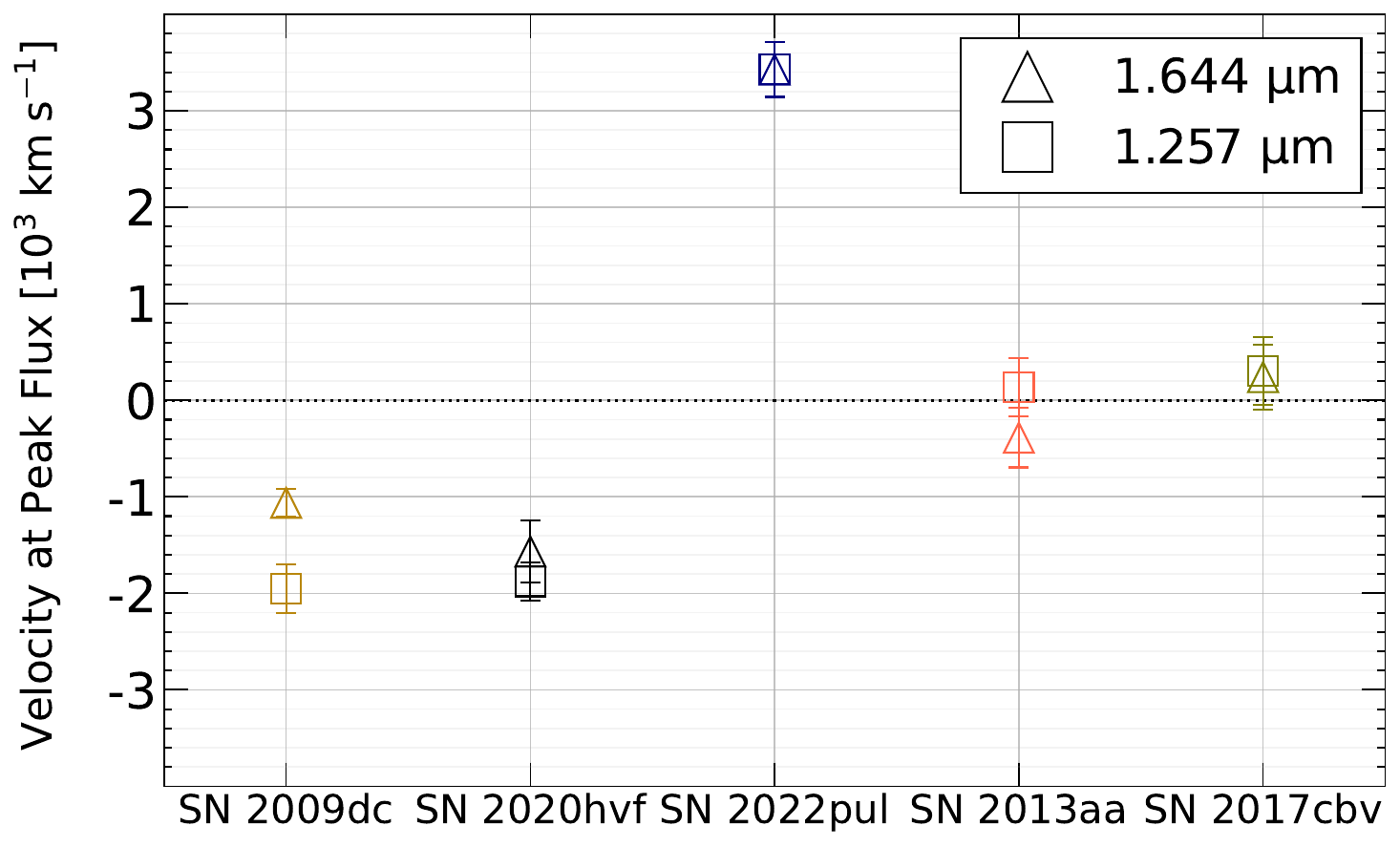}
    \caption{Velocity at peak flux of the [Fe~II] 1.257 and 1.644~\micron~features of each SN, which are measured via the procedure described in Section~\ref{subsec:analytical_peaks}. 03fg-like SNe have peaks that are non-zero but consistent between SNe, whereas the normal SNe velocities at peak flux are roughly consistent with $0~\mr{km~s^{-1}}$. The peak values plotted here are tabulated in Table \ref{tab:tiltresults}.}
    \label{fig:vpeaks}
\end{figure}

Figure~\ref{fig:vpeaks} presents the results of this fitting procedure for all three of our 03fg-like SNe, plus SNe~2013aa and 2017cbv. From this plot, velocity at peak flux seems to vary  over a range of~-2000~km~s$^{-1}$ to 3000~km~s$^{-1}$ within the sample of SNe.
Furthermore, there is strong agreement between velocities at peak flux for the 1.257 and 1.644~$\micron$ features for each 03fg-like SN. 
The fact that these peaks are simultaneously blue-/red-shifted the same way for each 03fg-like SN we analyze suggests that chemical asymmetries within the Fe-rich core of the ejecta may be a common characteristic of 03fg-like SNe. Conversely, our values for velocity at peak flux in the normal SNe~Ia are not asymmetric, as they are close to $\sim$0~$\mr{km~s^{-1}}$, demonstrating that the profile asymmetries seen in 03fg-like SNe are distinct.

\begin{deluxetable}{ccc}
  \tablecaption{Measurements of velocity at peak flux for asymmetric features.\label{tab:tiltresults}} 
  \tablehead{ \colhead{SN} & \colhead{1.257~$\micron$} & \colhead{1.644~$\micron$}}
  \startdata
    SN~2009dc &$-$1940 (240)&$-$1080 (150)\\
    SN~2013aa &120 (300)&$-$400 (310)\\
    SN~2017cbv&310 (340)&240 (340)\\
    SN~2020hvf&$-$1890 (140)&$-$1562 (320)\\
    SN~2022pul&3455 (290)&3448 (280)\\
    \tableline
  \enddata
  \tablecomments{Velocities at peak flux for the 1.257 and 1.644~$\micron$ features, given in $\kmps$, with uncertainties in parentheses. The methods used in obtaining these values are detailed in Section~\ref{subsec:analytical_peaks}.}
\end{deluxetable}


\subsection{Profile Tilts} \label{subsec:analytical_tilts}
The 03fg-like SNe in the sample also have varying degrees of asymmetry, or tilt, across the spectral regions of interest. Quantifying this tilt may provide important information about the chemical distributions of 03fg-like SNe, as well as provide a useful parameter for future observations. We hereafter refer to tilt symbolically as $m_{T}$. To quantify $m_{T}$, we use a Monte Carlo (MC) fitting method, where we calculate the slope of a straight line fitted to the skewed top portion of the feature. The sign and magnitude of the slope respectively describe the direction and steepness of $m_{T}$, where we define positive $m_{T}$ to be a red-peaked slope.

Each MC iteration is constructed using the same method described in Section~\ref{subsec:analytical_peaks}. However, for each iteration, we fit the tilted top of the perturbed smooth profile with a straight line, where the bounds on this linear fit are manually defined to be the locations of the two outer edges of the tilted top of the feature. We call the velocity separation between these two bounds $\Delta v$, and we determine its uncertainty the same way as the velocity at peak flux. We conduct 500 iterations of this process, where the mean and standard deviation of the sample of slopes taken from each linear fit are our measurement of $m_T$ in the 1.257 or 1.644~$\micron$ line.

\begin{figure}
    \centering
    \includegraphics[width=\linewidth]{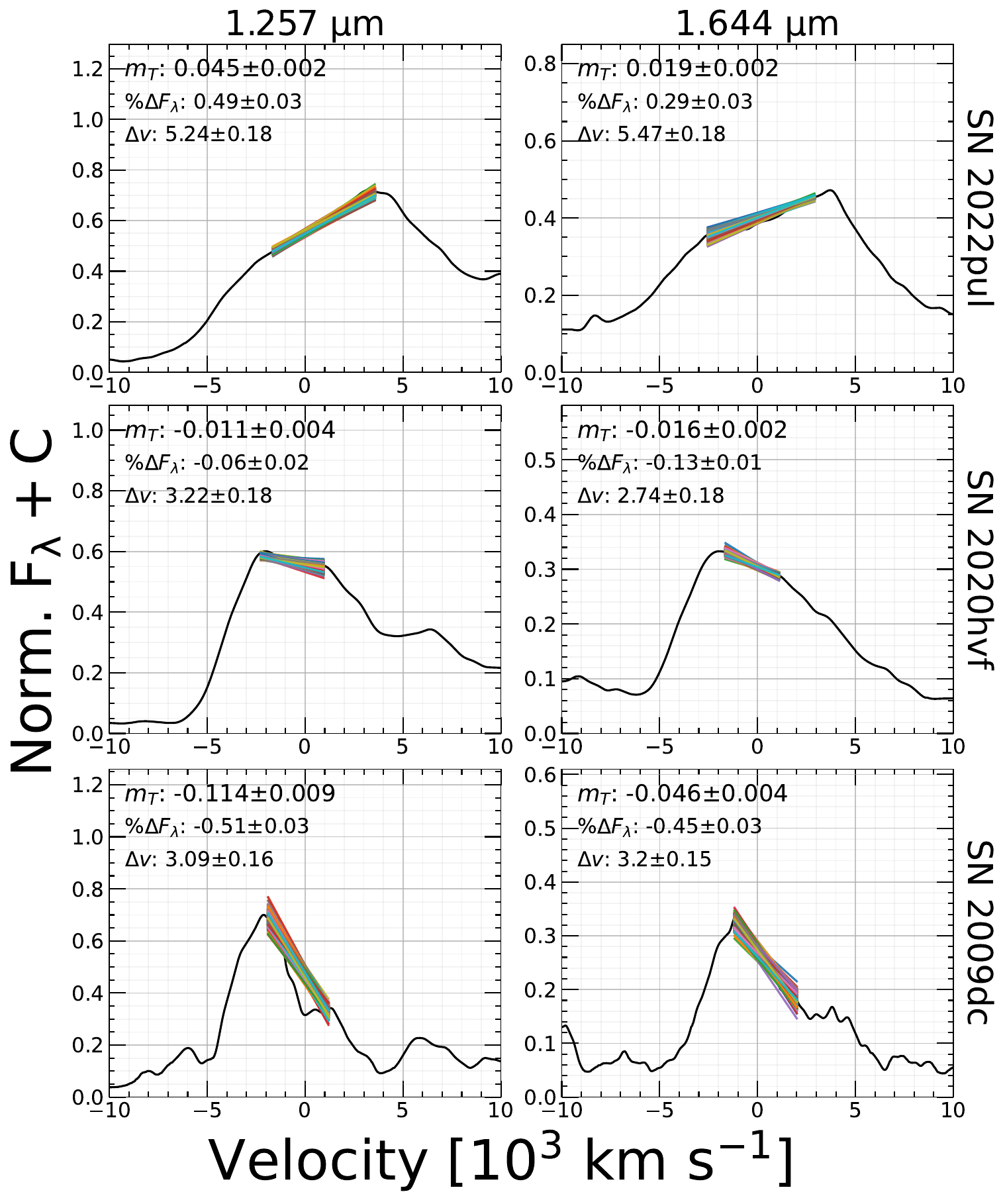}
    \caption{Visualization of the results of the MC tilt measurements described in Section~\ref{subsec:analytical_tilts}. Individual spectra are replaced with a 1D Gaussian-filtered smooth curve with smoothing length of 9 pixels to improve the visibility of each MC line fit across the tilt of each feature.}
    \label{fig:tilts_megaplot}
\end{figure}

In addition, we choose to examine the percent-change in flux relative to feature maximum $\%\Delta F_\lambda$ across $\Delta v$ in order to quantitatively compare the 1.257 and 1.644~$\micron$ regions. Our measurements of $m_T$, $\Delta v$, and $\%\Delta F_\lambda$ are listed in Table~\ref{tab:tiltresults} and shown graphically in Fig.~\ref{fig:tilts_megaplot}.

\begin{figure}
    \centering
    \includegraphics[width=\linewidth]{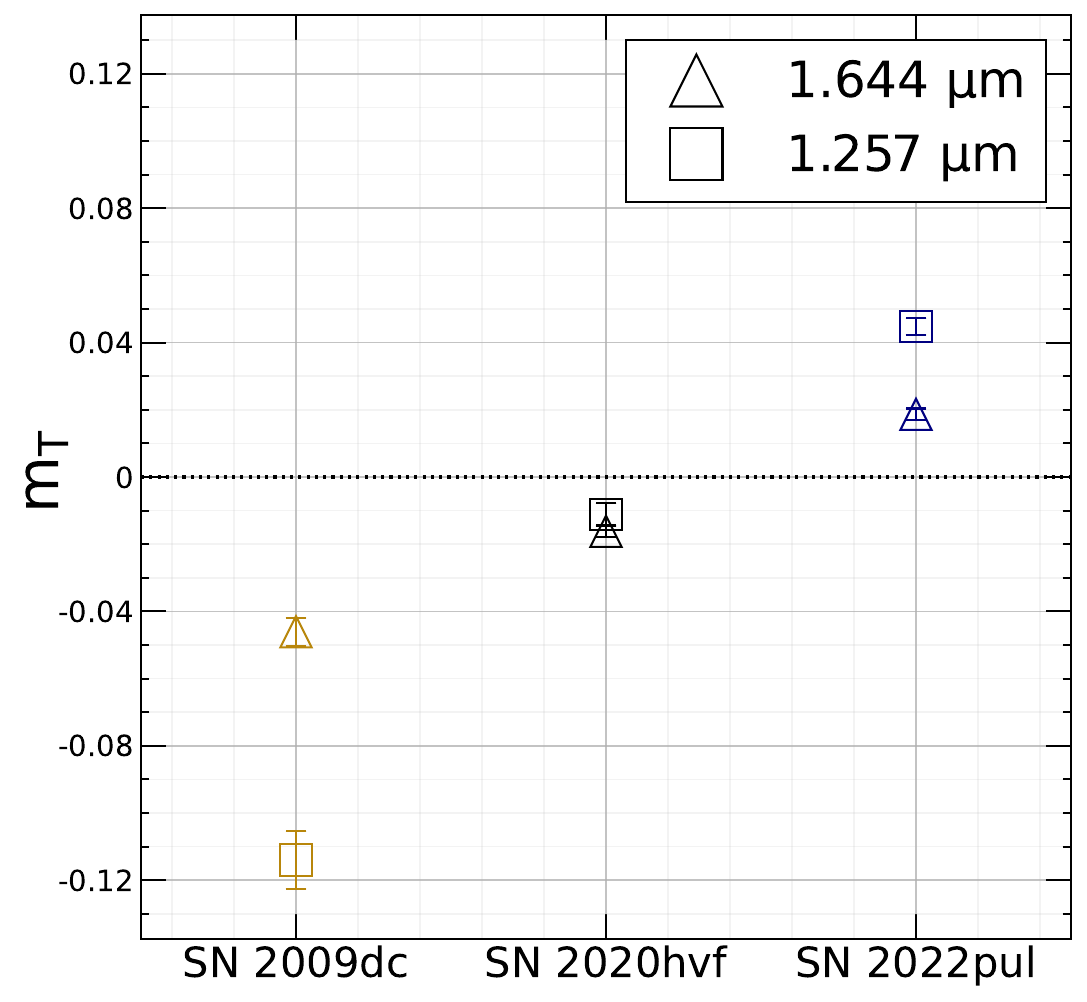}
    \includegraphics[width=1.02\linewidth]{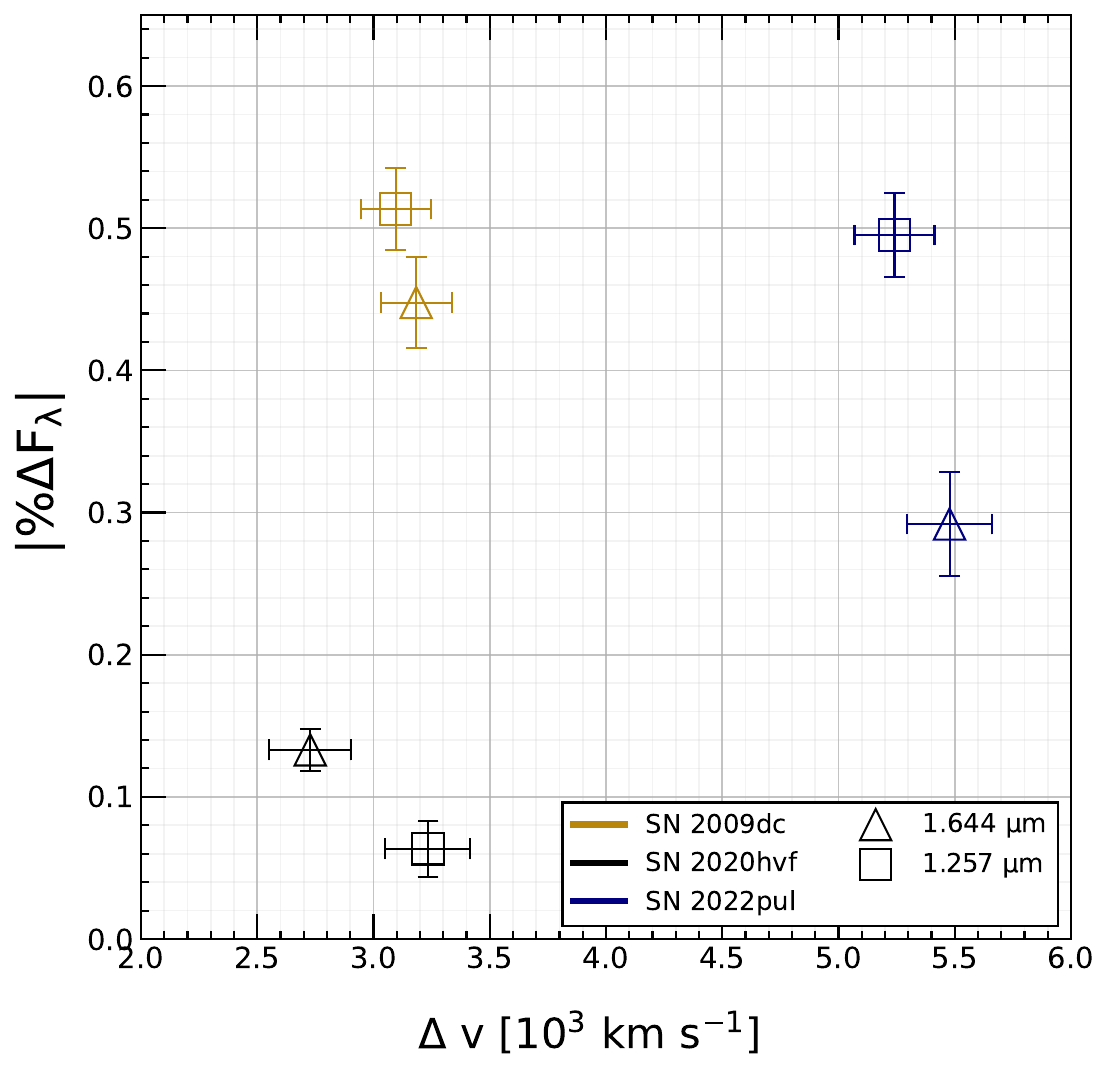}
    \caption{\textit{Top:} Tilt $m_T$ of the [Fe~II] 1.257 and 1.644$~\micron$ features of each 03fg-like SN, which are measured via the procedure described in Section~\ref{subsec:analytical_tilts}. The tilts are consistent across features for SN~2020hvf and mostly consistent for SN~2022pul, but not for SN~2009dc, the reasons for which are discussed in Section~\ref{subsec:analytical_tilts} (See also Table~\ref{tab:tiltresults} and Fig.~\ref{fig:tilts_megaplot}). \textit{Bottom:} Comparison of absolute $\%\Delta F_\lambda$ and $\Delta v$, which are measured by the procedures explained in Section~\ref{subsec:analytical_tilts}. No correlation is observable here, suggesting that asymmetric features have varied changes in flux between their edges, regardless of $\Delta v$.}
    \label{fig:tilts_03fglikes}
\end{figure}

The top panel of Fig.~\ref{fig:tilts_03fglikes} plots the values of $m_T$ for each feature in each 03fg-like SN. We observe an agreement in tilt between the 1.257 and 1.644~$\micron$ features in SN~2020hvf and SN~2022pul, but a large disparity in $m_T$ exists between these features in SN~2009dc. This disparity may be attributed to spectral line blending in SN~2009dc at about $4000~\mr{km~s^{-1}}$ redwards of 1.644~\micron~that artificially ``raises" the red edge of the MC line fits, resulting in a positive skew for $m_T$. Conversely, the peak of the 1.257~\micron~feature in SN~2009dc may be unusually strong, resulting in $m_T$ being more negative.

The bottom panel of Fig.~\ref{fig:tilts_03fglikes} plots the absolute percentage change in flux against the velocity separation between the outer edges of the tilted top for each SN. 
The absolute percentage change in flux is similar between the 1.257 and 1.644~$\micron$ features for each 03fg-like SN, with the exception of SN~2022pul. 
Generally, no correlation between $|\%\Delta F_\lambda|$ and $\Delta v$ is observed between SNe. This implies that, for tilted features with similar widths, a range of changes in flux across the tilt are possible. However, the overall consistency in tilting across features for each SN suggests that these profiles are not dominated by spherical [Fe~II] emission, but rather asymmetric [Fe~II] emission from an aspherical chemical distribution in the core of the ejecta. Future detailed NLTE spectral modeling is encouraged to corroborate this further.

\begin{deluxetable*}{c|ccc|ccc} 
  \tablecaption{Measurements of asymmetric feature morphology.\label{tab:tiltresults}} 
  \tablehead{\colhead{} & \colhead{} & \colhead{1.257~\micron} & \colhead{} & \colhead{} & \colhead{1.644~\micron} & \colhead{} \\ \tableline \colhead{SN} & \colhead{$m_T$} & \colhead{$\%\Delta F_\lambda$} & \colhead{$\frac{\Delta v}{\mr{10^3~km~s^{-1}}}$} & \colhead{$m_T$} & \colhead{$\%\Delta F_\lambda$} & \colhead{$\frac{\Delta v}{\mr{10^3~km~s^{-1}}}$} }
  \startdata
    SN~2009dc & -0.114 (0.009)& -0.51 (0.03)& 3.09 (0.16) & -0.046 (0.004)& -0.45 (0.03)& 3.20 (0.15)\\
    SN~2020hvf& -0.011 (0.004)& -0.06 (0.02)& 3.22 (0.18) & -0.016 (0.002)& -0.13 (0.01)& 2.74 (0.18)\\
    SN~2022pul& 0.045 (0.002)& 0.49 (0.03)& 5.24 (0.18) & 0.019 (0.002)& 0.29 (0.03)& 5.47 (0.18)
  \enddata
  \tablecomments{Tilt $m_T$ is the degree to which the asymmetry in the feature tilts; percent change in flux $\%\Delta F_\lambda$ is the ratio of the flux at the red peak and the flux at the blue peak of the asymmetric feature; velocity separation $\Delta v$ is the difference in velocity space between the red and blue peaks of the feature. Uncertainties for these values are recorded in parentheses; the methods used in obtaining them are detailed in Section~\ref{subsec:analytical_tilts}.}
\end{deluxetable*}

\subsection{Residual-Testing 03fg-like Features Against Normal SNe Ia} \label{subsec:analytical_residuals}
\begin{figure*}
    \includegraphics[width=\textwidth]{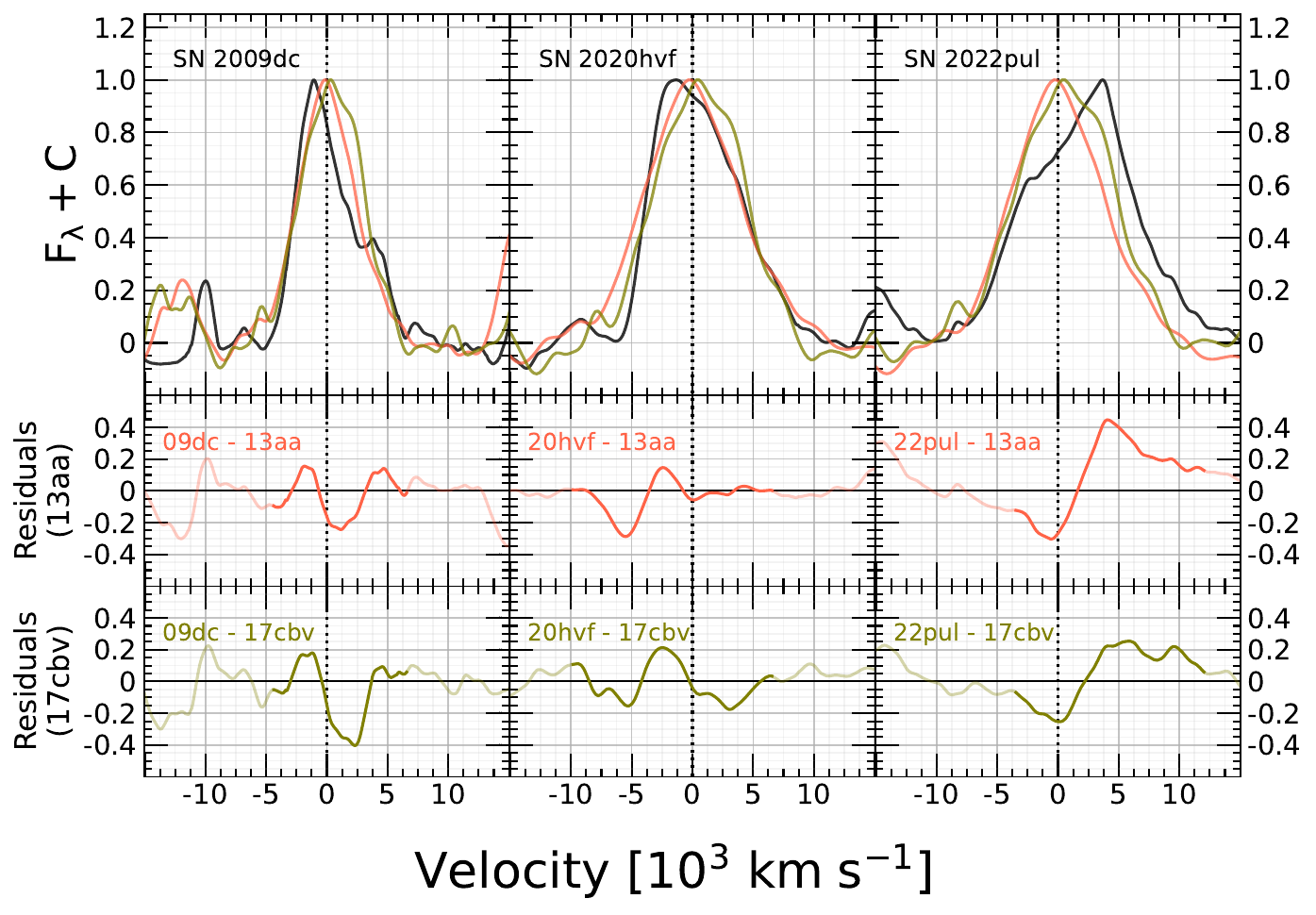}
    \caption{Residual plots of the 1.644~\micron~features of SN~2009dc, SN~2020hvf, and SN~2022pul (black) against those of both SN~2013aa (red) and SN~2017cbv (green). The procedure for obtaining these residual plots is detailed in Section~\ref{subsec:analytical_residuals}. Highlighted regions in the bottom two rows represent the most significant deviations in profile shape for each 03fg-like SN.}
    \label{fig:residuals}
\end{figure*}

To further differentiate the line profiles of 03fg-like SNe from those of normal SNe, we examine the residuals between the 1.644~$\micron$ features of 03fg-likes and standard SNe~Ia. This allows us to graphically demonstrate how the 1.644$~\micron$ features of 03fg-like SNe deviate from a symmetric profile, and thus how their ejecta deviate from a spherical chemical distribution. Residual testing can also provide useful hints on what features may exist in 03fg-likes that distinguish them from normal SNe~Ia; however, we limit this analysis to the 1.644~$\micron$ feature since it is symmetric and exhibits minimal blending in the normal SNe.

1-D Gaussian filtering is again used to smooth each 03fg-like and normal SN spectrum. In preparing each residual test, the smoothed normal SNe  spectra are each scaled by eye such that the tails of the CN 1.644~\micron~lines match up with those of the 03fg-like they are being compared to; this is done without shifting the spectrum itself horizontally. All smoothed spectra are then normalized to 1 at peak flux, leaving two normal 1.644~\micron~features and one 03fg-like 1.644~\micron~feature superimposed with each other.

As such, for residual testing to be possible computationally, the normal SN spectra are interpolated to have the same resolution as the 03fg-like spectra. This way, direct calculations of residuals are possible.

Figure~\ref{fig:residuals} shows the results of the residual tests of each 03fg-like SN against SNe~2013aa and 2017cbv. For each 03fg-like SN, its residual against SN~2013aa is similar to the residual against SN~2017cbv. The residual pattern varies widely with each 03fg-like SN, however. SN~2009dc has a significant negative residual just redwards of 1.644~$\micron$ produced by the large tilt on the feature, with a minimum at 2000~km~s$^{-1}$. SN~2020hvf aligns somewhat closely with the normal SNe on its red side, but exhibits a steeper tilt on its blue side, leading to a minimum flux at -5500~km~s$^{-1}$. SN~2022pul has a very large negative residual centered at 0~km\,s$^{-1}$, and an even larger positive residual across its entire red side peaking between 4000-6000~km\,s$^{-1}$. 

Overall, there is no pattern to the residual shapes of SN~2009dc, SN~2020hvf, and SN~2022pul. This suggests that there may be a continuum of viewing angle effects and/or shapes of asymmetric chemical distributions in the cores of 03fg-like SNe ejecta. Further residual testing of a larger sample of 03fg-like SNe is thus encouraged, especially considering its ability to visualize the full profile rather than just its peak or tilt.

\subsection{Velocity Fitting} \label{subsec:analytical_gaussian}
To quantify line asymmetries further, we fit the profiles in the 1.26 and 1.64$~\mr{\micron}$ regions with analytical functions, which has been done for other SNe in the nebular phase \citep[e.g.][]{Maguire_2018_126,Tucker20,Derkacy_2023_21aefx,Ashall24_JWST}. We note that this is likely an oversimplification of the problem, as it assumes that the features are produced from multiple symmetric line profiles, whereas 03fg-likes are likely to be composed of asymmetric chemical distributions that give rise to inherently non-symmetric line profiles (for observational examples of this, see \citealt{Hoeflich_2021_20qxp} and \citealt{Kwok_2024_22pul}). Nonetheless, we still choose to provide these fits for consistency, as it is especially applicable for analyses of spectral features with minimal blending~\citep{Diamond_2018_14j_164,Maguire_2018_126,kumar_2023_13aa17cbv,Siebert_2023_asymmetricdd}.

We fit the data in the two regions of interest (1.257 and 1.644~$\mr{\micron}$) under the assumption that the [Fe~II] 1.257 and 1.644~$\micron$ transitions can both be characterized by two separate emitting regions that are each approximately Gaussian. We make the same assumption for emissions from other byproducts of $^{56}$Ni, such as Co, but not necessarily for stable Ni itself. These weaker lines which contribute to the flux in these regions are based on our line identifications from Section~\ref{sec:line_ids}.

The fits themselves follow a procedure similar to those described in~\cite{Diamond_2018_14j_164},~\cite{Maguire_2018_126},~\cite{Siebert_2020_19yvq_gaussians},~\cite{kumar_2023_13aa17cbv}, and~\cite{Siebert_2023_asymmetricdd}. The fitting process begins by subtracting a manually-bounded linear continuum from the feature. We then fit Gaussians to the continuum-subtracted feature with the \textsc{python} tool \textsc{curve\_fit()} from the \textsc{scipy.optimize} library~\citep{Jones_2001_scipy,Fadillah_2021_leastsq}. This tool produces optimal parameters and covariances on these parameters for each Gaussian component of the fit. The error on each parameter is then taken to be the square root of its covariance with itself. We allow up to 1000~$\chi^2$ iterations, where each Gaussian can vary in width, strength, and peak position independently of the ion producing the transition.
We then convert to velocity space the best-fit parameters for mean (hereafter $v_{\mr{peak}}$) and standard deviation (which is also converted to FWHM and henceforth called $v_{\mr{width}}$).

\begin{figure}[t]
    \centering
    \includegraphics[width=\linewidth]{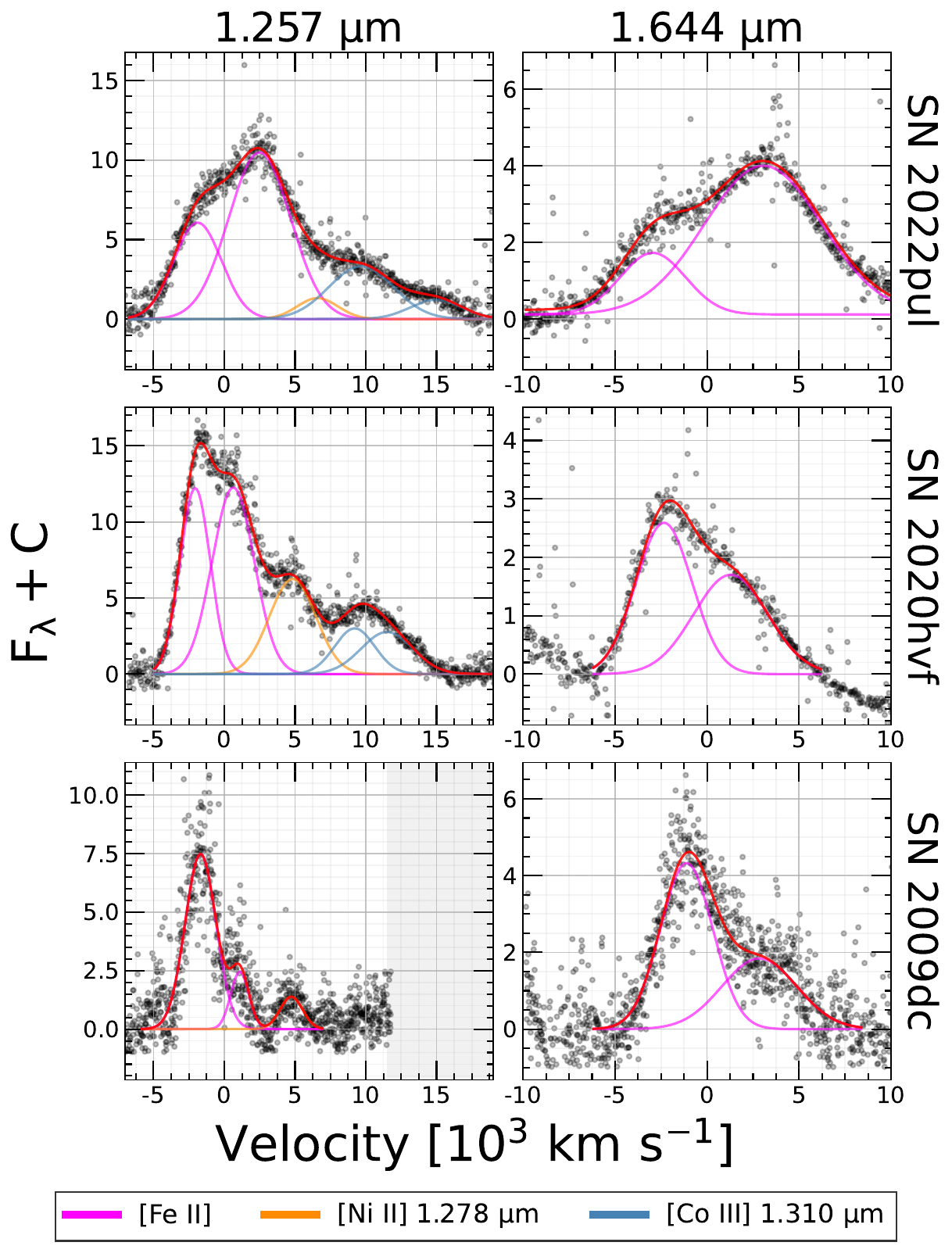}
    \caption{Gaussian curve fitting results for the 1.644~$\micron$ and 1.257~$\micron$ features of our 03fg-like SNe. The gray region past $\sim11000\mr{~km~s^{-1}}$ for the 1.257~\micron~feature of SN~2009dc is a telluric region left unobserved by the XShooter instrument (Table~\ref{tab:obs}).}
    \label{fig:gaussianfit_results}
\end{figure}

The graphical results of our Gaussian fits to the [Fe~II] 1.644~$\micron$ features are displayed in the right column of Fig.~\ref{fig:gaussianfit_results}. Each fit recreates its corresponding profile. The component with the largest amplitude also aligns with the peaked side of the feature, which is the blue side for SN~2009dc and SN~2020hvf and the red side for SN~2022pul.

The left column of Fig.~\ref{fig:gaussianfit_results} shows the results of our Gaussian fits to the [Fe~II] 1.257~$\micron$ features. In contrast to the 1.644~$\micron$ feature, there is blending around this complex; furthermore, the features that are blended vary between each 03fg-like SN. This results in different ionic species being identified for different SNe in the 1.257~$\micron$ region. For instance, every 03fg-like SN is fitted with a singular [Ni~II] 1.278~$\micron$ line just redwards of the ubiquitous [Fe~II] 1.257~$\micron$ lines; however, only SN~2020hvf and SN~2022pul are fitted with two [Co~III] 1.310~$\micron$ lines. In the case of a singular [Ni~II] component, which produces the best fits, this is physically permissible, but we cannot rule out double [Ni~II] emissions.

\begin{figure}[t]
    \centering
    \includegraphics[width=\linewidth]{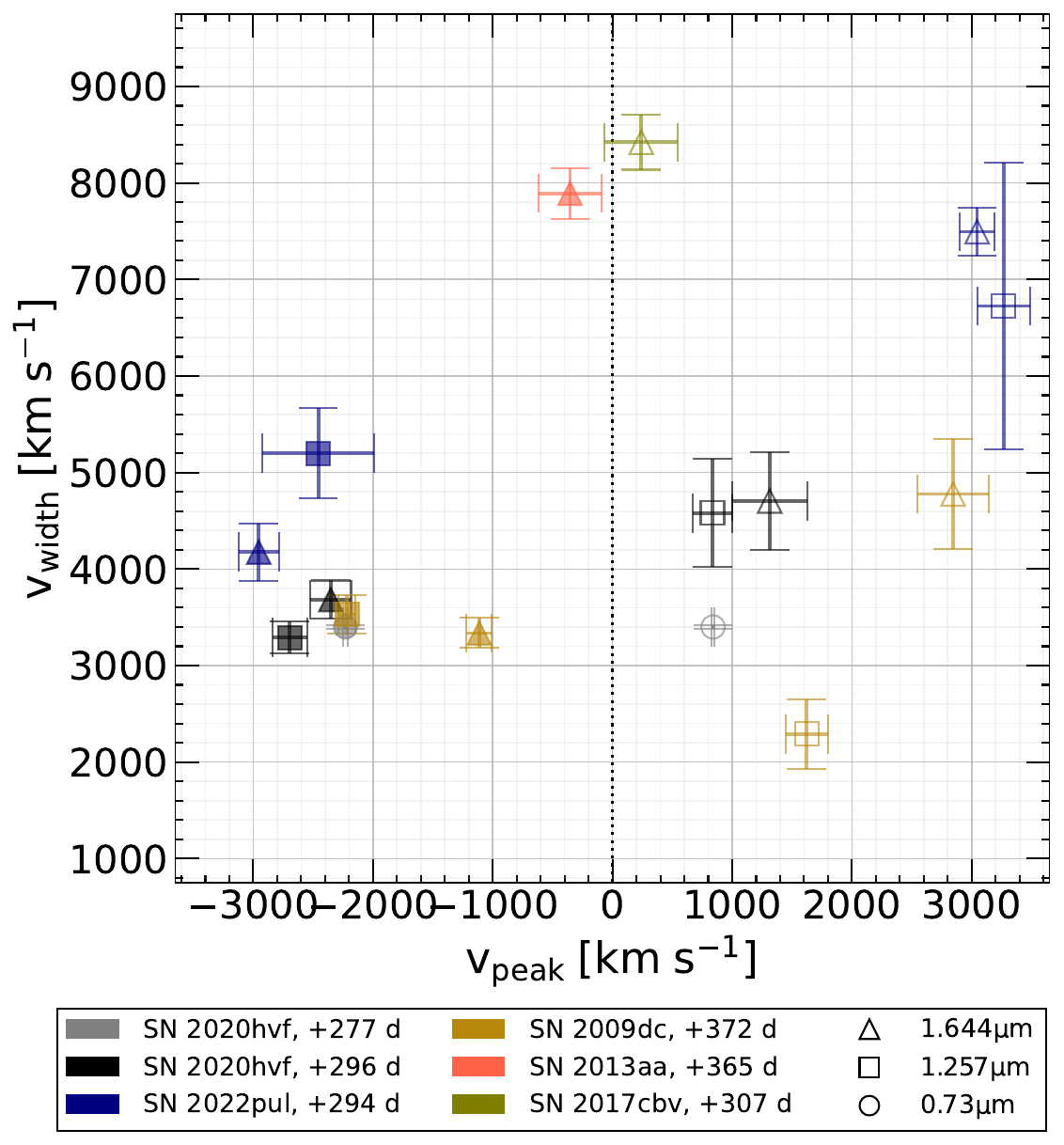}
    \caption{$v_{\mr{width}}$ plotted against $v_{\mr{peak}}$ for each Gaussian corresponding to the [Fe~II] 1.257 or 1.644~$\micron$ feature profile of each SN. Fit results for SNe 2013aa and 2017cbv are obtained from~\cite{kumar_2023_13aa17cbv}, while fit results for the [Fe~II] 0.73~$\micron$ feature of SN~2020hvf are from~\cite{Siebert_2023_asymmetricdd}. All measurements are made using a procedure similar to that described in Section~\ref{subsec:analytical_gaussian}.  The consistent separations in velocity between coupled components of the [Fe~II] features of the 03fg-likes are indicative of asymmetric chemical distributions in the cores of these explosions. No significant velocity offset, however, is observed in the [Fe~II] 1.644~$\micron$ features of the normal SNe.}
    \label{fig:aspherplot}
\end{figure}

The numerical results for $v_{\mr{peak}}$ and $v_{\mr{width}}$ from our Gaussian fits are shown in Fig.~\ref{fig:aspherplot}, which plots $v_{\mr{peak}}$ against $v_{\mr{width}}$ for each Gaussian representing [Fe~II] 1.257 or 1.644$~\micron$ emissions for each 03fg-like SN, along with those of each 1.644~$\micron$ component for each normal SN. For clarity, we define ``coupled" components to be the red and blue members of a double Gaussian fit corresponding to a singular spectral feature that is tilted. While it may seem natural to explain such coupled components as arising from two separate emitting spheres (i.e. a merger of two WDs), we stress here that this may not be the true physical composition of the ejecta. Multiple Gaussian emissions may still be an analytical way of explaining line profiles occurring from asymmetric chemical distributions. 
Only through detailed modeling, however, can this issue be disentangled. Regardless of their true nature, these coupled components allow us to examine the nonspherical nature of the emitting region. We report that in each of our 03fg-like SNe the difference in $v_{\mr{peak}}$ between coupled components is consistent in both the 1.257 and 1.644~$\micron$ features. 

For SN~2020hvf the blue components center around $\sim-2500~\mr{km~s^{-1}}$, and the red components $\sim$1000~$\mr{km~s^{-1}}$, making a separation of $\sim3500\mr{km~s^{-1}}$. SN~2022pul has a larger separation of 
$\sim$6000~$\mr{km~s^{-1}}$, with the blue components centered around $\sim$-3000~$\mr{km~s^{-1}}$ and the red components $\sim$3000~$\mr{km~s^{-1}}$. The split for SN~2009dc is more uncertain due to the low SNR of the spectrum, which makes it difficult to properly fit the weaker lines in the feature. Nonetheless, the split between the features is approximately $\sim$4000~$\mr{km~s^{-1}}$.

As~\cite{Siebert_2023_asymmetricdd} performed a similar procedure for fitting the optical Fe/Ni/Ca complex at $\mr{\sim7300\AA}$ for SN~2020hvf, we discuss how our results compare. One difference between the analysis of \cite{Siebert_2023_asymmetricdd} and our work is that \cite{Siebert_2023_asymmetricdd} add the additional constraint that the values of $v_{\mr{width}}$ are the same between coupled [Fe~II] 0.7155~$\mr{\micron}$ components. This is because the optical lines are much more blended than the NIR lines, so the relative widths cannot feasibly be determined. \cite{Siebert_2023_asymmetricdd} obtained a $\sim3000~\mr{km~s^{-1}}$ velocity separation between coupled [Fe~II] 0.7155~$\mr{\micron}$ components at a phase of +277~d, which agrees well with our results for the velocity separation between coupled components of [Fe~II] 1.257 and 1.644~$\mr{\micron}$ at +294~d. However, since our methods do not constrain $v_{\mr{width}}$ to be uniform between coupled components, we obtain $v_{\mr{width}}=\sim3500~\kmps$ for the blue components and $\sim4500~\kmps$ for the red components. Furthermore, \cite{Siebert_2023_asymmetricdd} reported a $\sim 2700~\mr{km~s^{-1}}$ separation between coupled [Ni~II] 0.7378~$\mr{\micron}$ emissions; in our data we find no strong need for a double Ni component in SN~2020hvf, but we reiterate that we cannot rule this out either. 

\subsection{Off-Center Delayed-Detonation Explosion Models} \label{subsec:analytical_ddt}
As discussed previously, one way to produce asymmetric line profiles in a SN~Ia explosion is through an off-center delayed detonation explosion \citep{Hoeflich_2021_20qxp}. In the context of 03fg-like SNe, this may occur in the core degenerate scenario. Previous modeling efforts on 03fg-like SNe have shown that a DDT may be a key component to explaining their peculiarities~\citep[e.g.][]{Hsiao_2020_carnegieii,Lu_2021_15hy}. In  off-center DDTs, the ignition starts near the center and the flame propagates as a deflagration before transitioning to a detonation at a specific mass coordinate. This off-center DDT can thus produce asymmetric chemical distributions and, by extension, off-center line profiles for various viewing angles \citep{2017ApJ...846...58H,Hoeflich_2021_20qxp,Derkacy_2023_21aefx,Ashall24_JWST}.

As it is outside the scope of this work to produce new models specific for 03fg-like SNe Ia, we visually compare the 1.644~$\micron$ lines of our 03fg-likes with DDT models created by~\cite{Hoeflich_2021_20qxp} in their analysis of the under-luminous type Ia SN~2020qxp, which also exhibits an asymmetric 1.644~$\micron$ line. These models demonstrate tilting in the 1.644~$\micron$ line due to an asymmetric chemical distribution produced by an off-center DDT observed from various viewing angles ($-90^\circ,\:-30^\circ,\:0^\circ,\:+30^\circ,\text{ and }+90^\circ$) relative to the equator (see~\citealt{Hoeflich_2021_20qxp} for more details). 
Although they are located in different areas of the luminosity width relation, both SN~2020qxp and 03fg-like SNe Ia have low ionization states relative to normal SNe Ia, and hence similar core structures \citep{Lu_2021_15hy,Hoeflich_2021_20qxp}. Thus, these models make a useful reference for testing this explosion scenario against the diverse range of data for our 03fg-like SNe. 

\begin{figure}[t]
    \centering
    \includegraphics[width=0.5\textwidth]{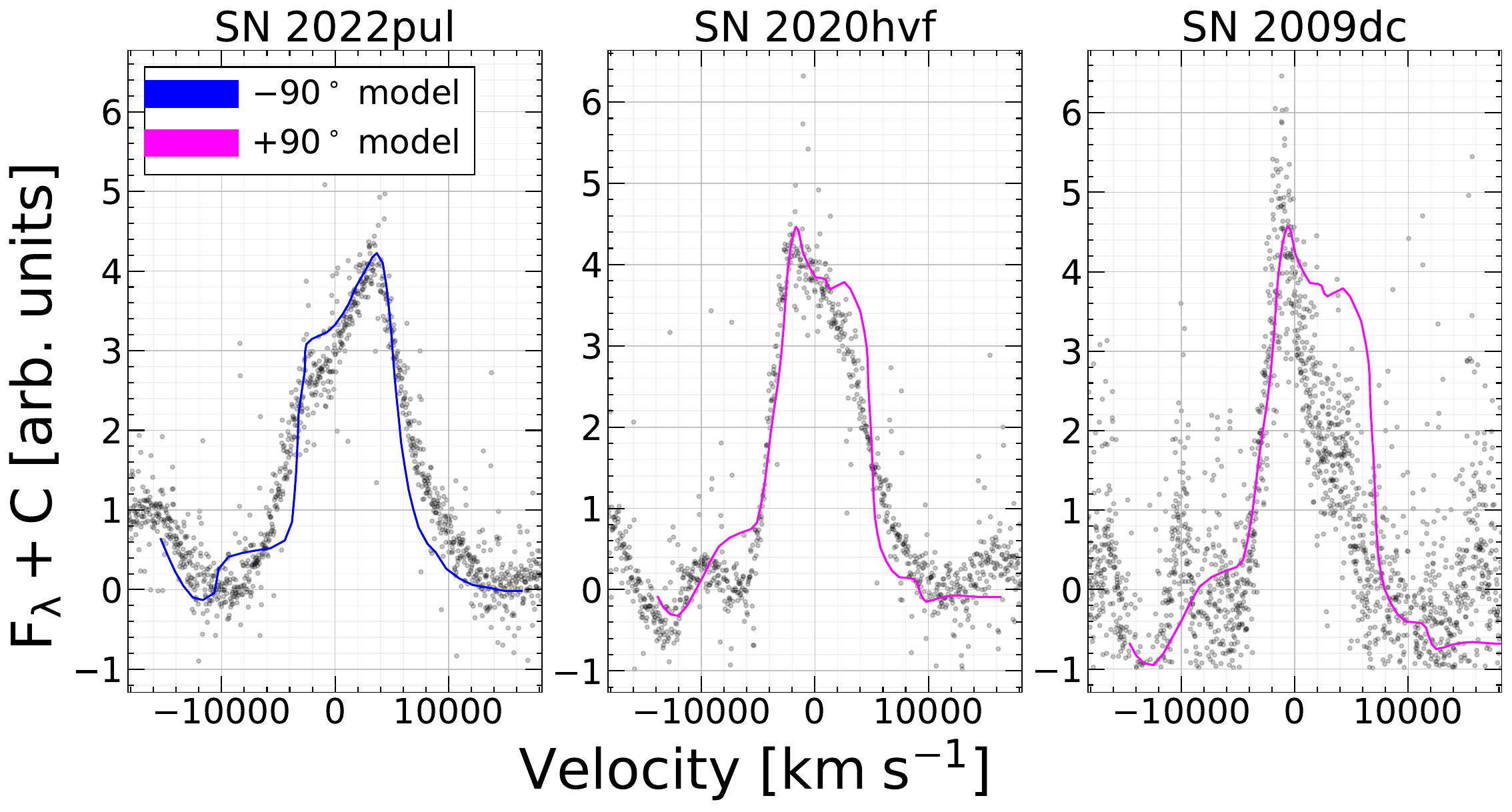}
    \caption{Comparison of off center DDT explosion models from~\cite{Hoeflich_2021_20qxp} overlaid with spectral observations of the 1.644~$\micron$ features of SN~2009dc, SN~2020hvf, and SN~2022pul. Note that the model spectra were adjusted in both flux scale and velocity space for comparison purpose, see the text for more details.}
    \label{fig:ddtfits}
\end{figure}

Despite having similar ionization states, SN~2020qxp and 03fg-like SNe 
have significant differences in luminosity, $^{56}$Ni mass, and Fe distributions. Hence, the model [Fe~II] $1.644~\micron$ lines have to be modified to align with those of the 03fg-like spectra in order to control for these variables. 
In short, we scale and shift the flux of each model vertically so that the peaks of the model and the data align in flux, controlling for luminosity; the widths of the models are also augmented to match the 03fg-likes, which controls for $^{56}$Ni mass (which is a strong predictor of [Fe~II] feature width). We overlay the DDT models with the 1.644~$\micron$ features of our 03fg-like SNe in Fig.~\ref{fig:ddtfits}. The choice of viewing angle for each DDT model being overlaid is based on its visual agreement with the tilt of the 03fg-like. 

There is good agreement between the $-90^\circ$ model and SN~2022pul, as well as good agreement between the $+90^\circ$ model and SN~2020hvf, supporting the idea that these objects may come from such an explosion scenario. On the other hand, for SN~2009dc, no model aside from $+90^\circ$ comes close to matching the profile shape. This may be a consequence of SN~2009dc having a different central density than SN~2020qxp, as the central density of the model determines the width of the tilted peak \citep{Hoeflich_2021_20qxp}. It is also possible that SN~2009dc is better explained by a completely separate explosion mechanism. 

Further modeling of off-center explosion models are encouraged for more robust model comparison tests of other peculiar SNe Ia like SN~2009dc.

\section{Conclusions} \label{sec:concl}
We present an analysis of three NIR spectra of 03fg-like SNe in the nebular phase. Before this work, only two such spectra existed in the literature~\citep{Taubenberger_2011_09dc,Siebert_2023_22pul}, where we contribute the third. While these nebular NIR spectra are a rare resource, we demonstrate that they are a powerful tool for determining the underlying physics in the explosion.
Specifically, we find that all three SNe in our sample show asymmetric line profiles in both the 1.26 and 1.64~$\mr{\micron}$ regions, where we identify [Fe II] 1.257 and 1.644~$\micron$ as the dominant lines contributing to each region. The shape and tilt of these line profiles are correlated within each SN, while there is a variation of tilt shape and width between different 03fg-like SNe. 

The analysis here suggests that asymmetric chemical distributions may be a common feature in 03fg-like SNe. We provide five quantitative methods for measuring the asymmetry and tilt in the spectra of the 03fg-like SNe. Through these methods, we show that 
\textit{i)} the peak velocities of both the 1.257 and 1.644~$\micron$ features vary between -2000 to +3000~km~s$^{-1}$ within our sample;
\textit{ii)} the tilts $m_T$ of the 1.257 and 1.644~$\micron$ features are consistent within 03fg-like SNe but can vary between them;
\textit{iii)} the residuals of 03fg-like SNe against normal SNe visually demonstrate a wide variety in the asymmetries of chemical distributions of 03fg-like SNe; 
\textit{iv)} multi-Gaussian fits demonstrate that, if spectral feature asymmetries are assumed to come from overlapping Gaussian emitting regions, the velocity separation of coupled components is consistent between features for each 03fg-like SN; and 
\textit{v)} comparison to pre-existing off-center DDT explosion models suggests that 03fg-like SNe produce  asymmetric chemical distributions which may be consistent with an off-center DDT scenario. 

These methods for measuring line profile asymmetry can be used to analyze future observations, even beyond the nebular-NIR domain. We must stress, though, that the process of multi-Gaussian fitting under the assumption of separate Gaussian emitting regions is not always physical when analyzing asymmetric chemical distributions (see Section~\ref{subsec:analytical_gaussian}). Additionally, we note that without detailed 3D NLTE modeling and a larger sample size of data, we cannot rule out that other spectral lines are contributing to the profiles in a way which it coincidentally appears that [Fe II] is the dominant ion in both the 1.257 and 1.644~$\micron$ regions.

Overall, there are two main models which could produce the line profiles presented in this work: the first is the merger of two WDs \citep[e.g.][]{Kwok_2024_22pul}, and the second is an off-center DDT explosion in the core degenerate scenario \cite[e.g.][]{Lu_2021_15hy,Hoeflich_2021_20qxp}. While it is outside the scope of this work to produce those models, we can speculate on what differs them. The global (not chemical) asymmetries between the models will differ (i.e. the core degenerate scenario is globally spherical, while the merger of two WDs is not). Only by combining early time continuum polarization measurements with medium-resolution late-time NIR spectra can these models be distinguished. Regardless of this, it is clear that significant advances in our understanding of the physics of the SNe~Ia can be made through analysis of NIR line profiles. Observationally speaking, however, the sample size is still small, and it is clear that additional medium-resolution NIR spectra are required to draw more robust conclusions. 

\begin{acknowledgments}
C.A, P.H., E.B., J.D., and K.M. acknowledge support by
NASA grants JWST-GO-02114, JWST-GO-02122, JWSTGO-03726, JWST-GO-04436, and JWST-GO-04522. C.A, P.H., E.B., J.D., and K.M. and M.S. 
acknowledge support by NASA grants JWST-GO-03726,
JWST-GO-04436, and JWST-GO-04522.
M.D.S. is funded by the Independent Research Fund Denmark (IRFD, grant number  10.46540/2032-00022B)
J.L. is supported by NSF-2206523 and DOE No. DE-SC0017955. 
L.G. acknowledges financial support from AGAUR, CSIC, MCIN and AEI 10.13039/501100011033 under projects PID2023-151307NB-I00, PIE 20215AT016, CEX2020-001058-M, and 2021-SGR-01270.

Some of the data presented here were obtained at Keck Observatory, which is a private 501(c)3 non-profit organization operated as a scientific partnership among the California Institute of Technology, the University of California, and the National Aeronautics and Space Administration. The Observatory was made possible by the generous financial support of the W. M. Keck Foundation.

The authors also wish to recognize and acknowledge the very significant cultural role and reverence that the summit of Mauna Kea has always had within the Native Hawaiian community. We are most fortunate to have the opportunity to conduct observations from this mountain.

\end{acknowledgments}

%

\vspace{5mm}
\facilities{Keck(NIRES), Magellan Baade(FIRE), VLT(XShooter), Las Cumbres Observatory}


\software{scipy~\citep{Jones_2001_scipy,Fadillah_2021_leastsq}}






\clearpage
\bibliographystyle{aasjournal}
\bibliography{sample631}

\begin{thebibliography}{}
\expandafter\ifx\csname natexlab\endcsname\relax\def\natexlab#1{#1}\fi
\providecommand{\url}[1]{\href{#1}{#1}}
\providecommand{\dodoi}[1]{doi:~\href{http://doi.org/#1}{\nolinkurl{#1}}}
\providecommand{\doeprint}[1]{\href{http://ascl.net/#1}{\nolinkurl{http://ascl.net/#1}}}
\providecommand{\doarXiv}[1]{\href{https://arxiv.org/abs/#1}{\nolinkurl{https://arxiv.org/abs/#1}}}

\bibitem[{{Ashall} {et~al.}(2021){Ashall}, {Lu}, {Hsiao}, {Hoeflich},
  {Phillips}, {Galbany}, {Burns}, {Contreras}, {Krisciunas}, {Morrell},
  {Stritzinger}, {Suntzeff}, {Taddia}, {Anais}, {Baron}, {Brown}, {Busta},
  {Campillay}, {Castell{\'o}n}, {Corco}, {Davis}, {Folatelli}, {F{\"o}rster},
  {Freedman}, {Gonzal{\'e}z}, {Hamuy}, {Holmbo}, {Kirshner}, {Kumar}, {Marion},
  {Mazzali}, {Morokuma}, {Nugent}, {Persson}, {Piro}, {Roth}, {Salgado},
  {Sand}, {Seron}, {Shahbandeh}, \& {Shappee}}]{Ashall_2021_03fglikes}
{Ashall}, C., {Lu}, J., {Hsiao}, E.~Y., {et~al.} 2021, \apj, 922, 205,
  \dodoi{10.3847/1538-4357/ac19ac}

\bibitem[{{Ashall} {et~al.}(2024){Ashall}, {Hoeflich}, {Baron}, {Shahbandeh},
  {DerKacy}, {Medler}, {Shappee}, {Tucker}, {Fereidouni}, {Mera}, {Andrews},
  {Baade}, {Bostroem}, {Brown}, {Burns}, {Burrow}, {Cikota}, {de Jaeger}, {Do},
  {Dong}, {Dominguez}, {Fox}, {Galbany}, {Hsiao}, {Krisciunas}, {Khaghani},
  {Kumar}, {Lu}, {Maund}, {Mazzali}, {Morrell}, {Patat}, {Pfeffer}, {Phillips},
  {Schmidt}, {Stangl}, {Stevens}, {Stritzinger}, {Suntzeff}, {Telesco}, {Wang},
  \& {Yang}}]{Ashall24_JWST}
{Ashall}, C., {Hoeflich}, P., {Baron}, E., {et~al.} 2024, \apj, 975, 203,
  \dodoi{10.3847/1538-4357/ad6608}

\bibitem[{Blondin {et~al.}(2023)Blondin, Dessart, Hillier, Ramsbottom, \&
  Storey}]{Blondin_2023_21aefx_models}
Blondin, S., Dessart, L., Hillier, D.~J., Ramsbottom, C.~A., \& Storey, P.~J.
  2023, Astronomy \& Astrophysics, 678, A170

\bibitem[{{Bose} {et~al.}(2025){Bose}, {Stritzinger}, {Ashall}, {Baron},
  {Hoeflich}, {Galbany}, {Hoogendam}, {Jensen}, {Kochanek}, {Post}, {Reguitti},
  {Elias-Rosa}, {Stanek}, {Lundqvist}, {Auchettl}, {Clocchiatti}, {Fiore},
  {Guti{\'e}rrez}, {Hinkle}, {Huber}, {de Jaeger}, {Pastorello}, {Payne},
  {Phillips}, {Shappee}, \& {Tucker}}]{Bose25}
{Bose}, S., {Stritzinger}, M.~D., {Ashall}, C., {et~al.} 2025, arXiv e-prints,
  arXiv:2501.04086, \dodoi{10.48550/arXiv.2501.04086}

\bibitem[{Childress {et~al.}(2011)Childress, Aldering, Aragon, Antilogus,
  Bailey, Baltay, Bongard, Buton, Canto, Chotard,
  {et~al.}}]{Childress_2011_metalpoor_07if}
Childress, M., Aldering, G., Aragon, C., {et~al.} 2011, The Astrophysical
  Journal, 733, 3

\bibitem[{{Cushing} {et~al.}(2004){Cushing}, {Vacca}, \&
  {Rayner}}]{Cushing_2004_Spextool}
{Cushing}, M.~C., {Vacca}, W.~D., \& {Rayner}, J.~T. 2004, \pasp, 116, 362,
  \dodoi{10.1086/382907}

\bibitem[{DerKacy {et~al.}(2023)DerKacy, Ashall, Hoeflich, Baron, Shappee,
  Baade, Andrews, Bostroem, Brown, Burns, {et~al.}}]{Derkacy_2023_21aefx}
DerKacy, J., Ashall, C., Hoeflich, P., {et~al.} 2023, The Astrophysical Journal
  Letters, 945, L2

\bibitem[{Diamond {et~al.}(2018)Diamond, Hoeflich, Hsiao, Sand, Sonneborn,
  Phillips, Hristov, Collins, Ashall, Marion, {et~al.}}]{Diamond_2018_14j_164}
Diamond, T., Hoeflich, P., Hsiao, E., {et~al.} 2018, The Astrophysical Journal,
  861, 119

\bibitem[{Diamond {et~al.}(2015)Diamond, Hoeflich, \&
  Gerardy}]{Diamond_2015_nir}
Diamond, T.~R., Hoeflich, P., \& Gerardy, C.~L. 2015, The Astrophysical
  Journal, 806, 107

\bibitem[{{Dimitriadis} {et~al.}(2023){Dimitriadis}, {Maguire}, {Karambelkar},
  {Lebron}, {Liu}, {Kozyreva}, {Miller}, {Ridden-Harper}, {Anderson}, {Chen},
  {Coughlin}, {Della Valle}, {Drake}, {Galbany}, {Gromadzki}, {Groom},
  {Guti{\'e}rrez}, {Ihanec}, {Inserra}, {Johansson}, {M{\"u}ller-Bravo},
  {Nicholl}, {Polin}, {Rusholme}, {Schulze}, {Sollerman}, {Srivastav},
  {Taggart}, {Wang}, {Yang}, \& {Young}}]{Dimitriadis_2023_21znydd}
{Dimitriadis}, G., {Maguire}, K., {Karambelkar}, V.~R., {et~al.} 2023, \mnras,
  521, 1162, \dodoi{10.1093/mnras/stad536}

\bibitem[{Fadillah {et~al.}(2021)Fadillah, Idrus, \&
  Hasan}]{Fadillah_2021_leastsq}
Fadillah, M. H. A.~Z., Idrus, B., \& Hasan, M.~K. 2021

\bibitem[{Filippenko {et~al.}(1992a)Filippenko, Richmond, Branch, Gaskell,
  Herbst, Ford, Treffers, Matheson, Ho, Dey,
  {et~al.}}]{Filippenko_1992_phot_91bg}
Filippenko, A.~V., Richmond, M.~W., Branch, D., {et~al.} 1992a, Astronomical
  Journal (ISSN 0004-6256), vol. 104, no. 4, p. 1543-1556, 1684., 104, 1543

\bibitem[{Filippenko {et~al.}(1992b)Filippenko, Richmond, Matheson, Shields,
  Burbidge, Cohen, Dickinson, Malkan, Nelson, Pietz,
  {et~al.}}]{Filippenko_1992_91t}
Filippenko, A.~V., Richmond, M.~W., Matheson, T., {et~al.} 1992b, Astrophysical
  Journal, Part 2-Letters (ISSN 0004-637X), vol. 384, Jan. 1, 1992, p.
  L15-L18., 384, L15

\bibitem[{Fink {et~al.}(2007)Fink, Hillebrandt, \&
  R{\"o}pke}]{Fink_2007_doubledet}
Fink, M., Hillebrandt, W., \& R{\"o}pke, F. 2007, Astronomy \& Astrophysics,
  476, 1133

\bibitem[{Ganeshalingam {et~al.}(2012)Ganeshalingam, Li, Filippenko, Silverman,
  Chornock, Foley, Matheson, Kirshner, Milne, Calkins,
  {et~al.}}]{Ganeshalingam_2012_02es}
Ganeshalingam, M., Li, W., Filippenko, A.~V., {et~al.} 2012, The Astrophysical
  Journal, 751, 142

\bibitem[{Hachinger {et~al.}(2012)Hachinger, Mazzali, Taubenberger, Fink,
  Pakmor, Hillebrandt, \& Seitenzahl}]{Hachinger_2012_coredegen}
Hachinger, S., Mazzali, P.~A., Taubenberger, S., {et~al.} 2012, Monthly Notices
  of the Royal Astronomical Society, 427, 2057

\bibitem[{{Hillebrandt} \& {Niemeyer}(2000)}]{HillebrandtNiemeyer_2000_sneia}
{Hillebrandt}, W., \& {Niemeyer}, J.~C. 2000, \araa, 38, 191,
  \dodoi{10.1146/annurev.astro.38.1.191}

\bibitem[{{Hoeflich} {et~al.}(2017){Hoeflich}, {Hsiao}, {Ashall}, {Burns},
  {Diamond}, {Phillips}, {Sand}, {Stritzinger}, {Suntzeff}, {Contreras},
  {Krisciunas}, {Morrell}, \& {Wang}}]{2017ApJ...846...58H}
{Hoeflich}, P., {Hsiao}, E.~Y., {Ashall}, C., {et~al.} 2017, \apj, 846, 58,
  \dodoi{10.3847/1538-4357/aa84b2}

\bibitem[{Hoeflich {et~al.}(2019)Hoeflich, Ashall, Fisher, Hristov, Collins,
  Hsiao, Wiedenhoever, Chakraborty, \& Diamond}]{Hoeflich_2019_multiwavelength}
Hoeflich, P., Ashall, C., Fisher, A., {et~al.} 2019, in Nuclei in the Cosmos
  XV, Springer, 187--194

\bibitem[{Hoeflich {et~al.}(2021)Hoeflich, Ashall, Bose, Baron, Stritzinger,
  Davis, Shahbandeh, Anand, Baade, Burns, {et~al.}}]{Hoeflich_2021_20qxp}
Hoeflich, P., Ashall, C., Bose, S., {et~al.} 2021, The Astrophysical Journal,
  922, 186

\bibitem[{Holcomb {et~al.}(2013)Holcomb, Guillochon, De~Colle, \&
  Ramirez-Ruiz}]{Holcomb_2013_hedet}
Holcomb, C., Guillochon, J., De~Colle, F., \& Ramirez-Ruiz, E. 2013, The
  Astrophysical Journal, 771, 14

\bibitem[{Hoogendam {et~al.}(2024)Hoogendam, Shappee, Brown, Tucker, Ashall, \&
  Piro}]{Hoogendam_2024_03fgUV}
Hoogendam, W., Shappee, B., Brown, P., {et~al.} 2024, The Astrophysical
  Journal, 966, 139

\bibitem[{Howell {et~al.}(2006)Howell, Sullivan, Nugent, Ellis, Conley,
  Le~Borgne, Carlberg, Guy, Balam, Basa, {et~al.}}]{Howell_2006_03fg}
Howell, A.~D., Sullivan, M., Nugent, P.~E., {et~al.} 2006, Nature, 443, 308

\bibitem[{Howell(2001)}]{Howell_2001_91bg}
Howell, D.~A. 2001, The Astrophysical Journal, 554, L193

\bibitem[{{Hoyle} \& {Fowler}(1960)}]{HoyleFowler_1960_sneia}
{Hoyle}, F., \& {Fowler}, W.~A. 1960, \apj, 132, 565, \dodoi{10.1086/146963}

\bibitem[{Hsiao {et~al.}(2020)Hsiao, Hoeflich, Ashall, Lu, Contreras, Burns,
  Phillips, Galbany, Anderson, Baltay, {et~al.}}]{Hsiao_2020_carnegieii}
Hsiao, E., Hoeflich, P., Ashall, C., {et~al.} 2020, The Astrophysical Journal,
  900, 140

\bibitem[{Iben~Jr \& Tutukov(1984)}]{IbenTutukov_1984_binaries_gws}
Iben~Jr, I., \& Tutukov, A.~V. 1984, Astrophysical Journal, Part 1 (ISSN
  0004-637X), vol. 284, Sept. 15, 1984, p. 719-744., 284, 719

\bibitem[{{Jerkstrand}(2017)}]{Jerkstrand17}
{Jerkstrand}, A. 2017, in Handbook of Supernovae, ed. A.~W. {Alsabti} \&
  P.~{Murdin}, 795, \dodoi{10.1007/978-3-319-21846-5_29}

\bibitem[{{Jiang} {et~al.}(2021){Jiang}, {Maeda}, {Kawabata}, {Doi},
  {Shigeyama}, {Tanaka}, {Tominaga}, {Nomoto}, {Niino}, {Sako}, {Ohsawa},
  {Schramm}, {Yamanaka}, {Kobayashi}, {Takahashi}, {Nakaoka}, {Kawabata},
  {Isogai}, {Aoki}, {Kondo}, {Mori}, {Arimatsu}, {Kasuga}, {Okumura},
  {Urakawa}, {Reichart}, {Taguchi}, {Arima}, {Beniyama}, {Uno}, \&
  {Hamada}}]{Jiang_2021_20hvf}
{Jiang}, J.-a., {Maeda}, K., {Kawabata}, M., {et~al.} 2021, \apjl, 923, L8,
  \dodoi{10.3847/2041-8213/ac375f}

\bibitem[{Jones {et~al.}(2001)Jones, Oliphant, {et~al.}}]{Jones_2001_scipy}
Jones, E., Oliphant, T., {et~al.} 2001

\bibitem[{Kashi \& Soker(2011)}]{KashiSoker_2011_coredegen}
Kashi, A., \& Soker, N. 2011, Monthly Notices of the Royal Astronomical
  Society, 417, 1466

\bibitem[{Khan {et~al.}(2011)Khan, Stanek, Stoll, \&
  Prieto}]{Khan_2011_metalpoor}
Khan, R., Stanek, K., Stoll, R., \& Prieto, J. 2011, The Astrophysical Journal
  Letters, 737, L24

\bibitem[{{Kumar}(2024)}]{Kumar24}
{Kumar}, S. 2024, in American Astronomical Society Meeting Abstracts, Vol. 243,
  American Astronomical Society Meeting Abstracts, 232.03

\bibitem[{Kumar {et~al.}(2023)Kumar, Hsiao, Ashall, Phillips, Morrell,
  Hoeflich, Burns, Galbany, Baron, Contreras, {et~al.}}]{kumar_2023_13aa17cbv}
Kumar, S., Hsiao, E.~Y., Ashall, C., {et~al.} 2023, The Astrophysical Journal,
  945, 27

\bibitem[{Kwok {et~al.}(2024)Kwok, Siebert, Johansson, Jha, Blondin, Dessart,
  Foley, Hillier, Larison, Pakmor, {et~al.}}]{Kwok_2024_22pul}
Kwok, L.~A., Siebert, M.~R., Johansson, J., {et~al.} 2024, The Astrophysical
  Journal, 966, 135

\bibitem[{Li {et~al.}(2001)Li, Filippenko, Treffers, Riess, Hu, \&
  Qiu}]{Li_2001_peculrate}
Li, W., Filippenko, A.~V., Treffers, R.~R., {et~al.} 2001, The Astrophysical
  Journal, 546, 734

\bibitem[{Li {et~al.}(2003)Li, Filippenko, Chornock, Berger, Berlind, Calkins,
  Challis, Fassnacht, Jha, Kirshner, {et~al.}}]{Li_2003_02cx}
Li, W., Filippenko, A.~V., Chornock, R., {et~al.} 2003, Publications of the
  Astronomical Society of the Pacific, 115, 453

\bibitem[{Liu {et~al.}(2017)Liu, Wang, Wu, \& Han}]{Liu_2017_cowd_herichwd}
Liu, D., Wang, B., Wu, C., \& Han, Z. 2017, Astronomy \& Astrophysics, 606,
  A136

\bibitem[{Liu {et~al.}(2025)Liu, Wang, Yang, Filippenko, Brink, Zheng, Zhang,
  Li, \& Yan}]{Liu_2025_CIin22pul}
Liu, J., Wang, X., Yang, Y., {et~al.} 2025, arXiv preprint arXiv:2502.18900

\bibitem[{Liu {et~al.}(2023)Liu, R{\"o}pke, \& Han}]{Liu_2023_iaexplosions}
Liu, Z.-W., R{\"o}pke, F.~K., \& Han, Z. 2023, Research in Astronomy and
  Astrophysics, 23, 082001

\bibitem[{{Livio} \& {Mazzali}(2018)}]{Livio18}
{Livio}, M., \& {Mazzali}, P. 2018, \physrep, 736, 1,
  \dodoi{10.1016/j.physrep.2018.02.002}

\bibitem[{Livio \& Riess(2003)}]{Livio_2003_iaprogenitors}
Livio, M., \& Riess, A.~G. 2003, The Astrophysical Journal, 594, L93

\bibitem[{{Lu} {et~al.}(2021){Lu}, {Ashall}, {Hsiao}, {Hoeflich}, {Galbany},
  {Baron}, {Phillips}, {Contreras}, {Burns}, {Suntzeff}, {Stritzinger},
  {Anais}, {Anderson}, {Brown}, {Busta}, {Castell{\'o}n}, {Davis}, {Diamond},
  {Falco}, {Gonzalez}, {Hamuy}, {Holmbo}, {Holoien}, {Krisciunas}, {Kirshner},
  {Kumar}, {Kuncarayakti}, {Marion}, {Morrell}, {Persson}, {Piro}, {Prieto},
  {Sand}, {Shahbandeh}, {Shappee}, \& {Taddia}}]{Lu_2021_15hy}
{Lu}, J., {Ashall}, C., {Hsiao}, E.~Y., {et~al.} 2021, \apj, 920, 107,
  \dodoi{10.3847/1538-4357/ac1606}

\bibitem[{Maeda {et~al.}(2023)Maeda, Jiang, Doi, Kawabata, \&
  Shigeyama}]{Maeda_2023_envelope_overluminous}
Maeda, K., Jiang, J.-a., Doi, M., Kawabata, M., \& Shigeyama, T. 2023, Monthly
  Notices of the Royal Astronomical Society, 521, 1897

\bibitem[{Maeda {et~al.}(2010)Maeda, Benetti, Stritzinger, R{\"o}pke,
  Folatelli, Sollerman, Taubenberger, Nomoto, Leloudas, Hamuy,
  {et~al.}}]{Maeda_2010_asymmetries}
Maeda, K., Benetti, S., Stritzinger, M., {et~al.} 2010, Nature, 466, 82

\bibitem[{Maguire {et~al.}(2018)Maguire, Sim, Shingles, Spyromilio, Jerkstrand,
  Sullivan, Chen, Cartier, Dimitriadis, Frohmaier, {et~al.}}]{Maguire_2018_126}
Maguire, K., Sim, S., Shingles, L., {et~al.} 2018, Monthly Notices of the Royal
  Astronomical Society, 477, 3567

\bibitem[{Marietta {et~al.}(2000)Marietta, Burrows, \&
  Fryxell}]{Marietta_2000_neb_progenitor}
Marietta, E., Burrows, A., \& Fryxell, B. 2000, The Astrophysical Journal
  Supplement Series, 128, 615

\bibitem[{Mazzali {et~al.}(1995)Mazzali, Danziger, \&
  Turatto}]{Mazzali_1995_91t}
Mazzali, P., Danziger, I., \& Turatto, M. 1995, Astronomy and Astrophysics, v.
  297, p. 509, 297, 509

\bibitem[{Mazzali {et~al.}(2007)Mazzali, Ropke, Benetti, \&
  Hillebrandt}]{Mazzali_2007_nebphase}
Mazzali, P.~A., Ropke, F.~K., Benetti, S., \& Hillebrandt, W. 2007, Science,
  315, 825

\bibitem[{Noebauer {et~al.}(2016)Noebauer, Taubenberger, Blinnikov, Sorokina,
  \& Hillebrandt}]{Noebauer_2016_csm}
Noebauer, U., Taubenberger, S., Blinnikov, S., Sorokina, E., \& Hillebrandt, W.
  2016, Monthly Notices of the Royal Astronomical Society, 463, 2972

\bibitem[{Pakmor {et~al.}(2011)Pakmor, Hachinger, R{\"o}pke, \&
  Hillebrandt}]{Pakmor_2011_violentsubluminous}
Pakmor, R., Hachinger, S., R{\"o}pke, F., \& Hillebrandt, W. 2011, Astronomy \&
  Astrophysics, 528, A117

\bibitem[{Pakmor {et~al.}(2010)Pakmor, Kromer, R{\"o}pke, Sim, Ruiter, \&
  Hillebrandt}]{Pakmor_2010_subluminousmergers}
Pakmor, R., Kromer, M., R{\"o}pke, F.~K., {et~al.} 2010, Nature, 463, 61

\bibitem[{Pakmor {et~al.}(2012)Pakmor, Kromer, Taubenberger, Sim, Röpke, \&
  Hillebrandt}]{Pakmor_2012_violentmergers}
Pakmor, R., Kromer, M., Taubenberger, S., {et~al.} 2012, The Astrophysical
  Journal Letters, 747, L10, \dodoi{10.1088/2041-8205/747/1/L10}

\bibitem[{Penney \& Hoeflich(2014)}]{Penney_2014_bfields}
Penney, R., \& Hoeflich, P. 2014, The Astrophysical Journal, 795, 84

\bibitem[{{Penney} \& {Hoeflich}(2014)}]{Penney14}
{Penney}, R., \& {Hoeflich}, P. 2014, \apj, 795, 84,
  \dodoi{10.1088/0004-637X/795/1/84}

\bibitem[{Siebert {et~al.}(2020)Siebert, Dimitriadis, Polin, \&
  Foley}]{Siebert_2020_19yvq_gaussians}
Siebert, M.~R., Dimitriadis, G., Polin, A., \& Foley, R.~J. 2020, The
  Astrophysical Journal Letters, 900, L27

\bibitem[{Siebert {et~al.}(2023)Siebert, Kwok, Johansson, Jha, Blondin,
  Dessart, Foley, Hillier, Larison, Pakmor, {et~al.}}]{Siebert_2023_22pul}
Siebert, M.~R., Kwok, L.~A., Johansson, J., {et~al.} 2023, The Astrophysical
  Journal, 960, 88

\bibitem[{{Siebert} {et~al.}(2023){Siebert}, {Foley}, {Zenati}, {Dimitriadis},
  {Schmidt}, {Yang}, {Davis}, {Taggart}, \&
  {Rojas-Bravo}}]{Siebert_2023_asymmetricdd}
{Siebert}, M.~R., {Foley}, R.~J., {Zenati}, Y., {et~al.} 2023, \apj, 958, 173,
  \dodoi{10.3847/1538-4357/ad037f}

\bibitem[{Silverman {et~al.}(2011)Silverman, Ganeshalingam, Li, Filippenko,
  Miller, \& Poznanski}]{Silverman_2011_09dc}
Silverman, J.~M., Ganeshalingam, M., Li, W., {et~al.} 2011, Monthly Notices of
  the Royal Astronomical Society, 410, 585

\bibitem[{Soker(2019)}]{Soker_2019_spherical}
Soker, N. 2019, New Astronomy Reviews, 87, 101535

\bibitem[{{Srivastav} {et~al.}(2023){Srivastav}, {Smartt}, {Huber},
  {Dimitriadis}, {Chambers}, {Fulton}, {Moore}, {Callan}, {Gillanders},
  {Maguire}, {Nicholl}, {Shingles}, {Sim}, {Smith}, {Anderson}, {de Boer},
  {Chen}, {Gao}, \& {Young}}]{Srivastav_2023_22ilv}
{Srivastav}, S., {Smartt}, S.~J., {Huber}, M.~E., {et~al.} 2023, \apjl, 943,
  L20, \dodoi{10.3847/2041-8213/acb2ce}

\bibitem[{Taubenberger {et~al.}(2011)Taubenberger, Benetti, Childress, Pakmor,
  Hachinger, Mazzali, Stanishev, Elias-Rosa, Agnoletto, Bufano,
  {et~al.}}]{Taubenberger_2011_09dc}
Taubenberger, S., Benetti, S., Childress, M., {et~al.} 2011, Monthly Notices of
  the Royal Astronomical Society, 412, 2735

\bibitem[{Taubenberger {et~al.}(2013)Taubenberger, Kromer, Hachinger, Mazzali,
  Benetti, Nugent, Scalzo, Pakmor, Stanishev, Spyromilio,
  {et~al.}}]{taubenberger_2013_nebular03fg}
Taubenberger, S., Kromer, M., Hachinger, S., {et~al.} 2013, Monthly Notices of
  the Royal Astronomical Society, 432, 3117

\bibitem[{Taubenberger {et~al.}(2019)Taubenberger, Floers, Vogl, Kromer,
  Spyromilio, Aldering, Antilogus, Bailey, Baltay, Bongard,
  {et~al.}}]{Taubenberger_2019_12dn}
Taubenberger, S., Floers, A., Vogl, C., {et~al.} 2019, Monthly Notices of the
  Royal Astronomical Society, 488, 5473

\bibitem[{{Tucker} {et~al.}(2020){Tucker}, {Shappee}, {Vallely}, {Stanek},
  {Prieto}, {Botyanszki}, {Kochanek}, {Anderson}, {Brown}, {Galbany},
  {Holoien}, {Hsiao}, {Kumar}, {Kuncarayakti}, {Morrell}, {Phillips},
  {Stritzinger}, \& {Thompson}}]{Tucker20}
{Tucker}, M.~A., {Shappee}, B.~J., {Vallely}, P.~J., {et~al.} 2020, \mnras,
  493, 1044, \dodoi{10.1093/mnras/stz3390}

\bibitem[{Wang \& Han(2012)}]{Wang_2012_iaprogenitors}
Wang, B., \& Han, Z. 2012, New Astronomy Reviews, 56, 122

\bibitem[{Webbink(1984)}]{Webbink_1984_doublewd}
Webbink, R. 1984, Astrophysical Journal, Part 1 (ISSN 0004-637X), vol. 277,
  Feb. 1, 1984, p. 355-360., 277, 355

\end{thebibliography}



\end{document}